\documentclass[sn-mathphys,Numbered]{sn-jnl}

\usepackage{graphicx}%
\usepackage{multirow}%
\usepackage{amsmath,amssymb,amsfonts}%
\usepackage{amsthm}%
\usepackage{mathrsfs}%
\usepackage[title]{appendix}%
\usepackage{xcolor}%
\usepackage{textcomp}%
\usepackage{manyfoot}%
\usepackage{booktabs}%
\usepackage{algorithm}%
\usepackage{algorithmicx}%
\usepackage{algpseudocode}%
\usepackage{listings}%
\usepackage{amsmath} 
\DeclareMathOperator*{\argmin}{arg\,min} 

\usepackage[utf8]{inputenc}
\DeclareUnicodeCharacter{200E}{}

\theoremstyle{thmstyleone}%
\theoremstyle{thmstyletwo}%

\theoremstyle{thmstylethree}%

\begin{document}

\title[The Functional Voting Classifier]{Supervised Learning via Ensembles of
Diverse Functional Representations: the Functional Voting Classifier}


\author[1]{\fnm{Donato} \sur{Riccio}}\email{donato.riccio@studenti.unicampania.it}

\author*[2]{\fnm{Fabrizio} \sur{Maturo}}\email{fabrizio.maturo@unimercatorum.it}

\author[3]{\fnm{Elvira} \sur{Romano}}\email{elvira.romano@unicampania.it}

\affil[1]{Machine Learning Engineer and Student in the Data Science Master's Degree Program of the University of Campania Luigi Vanvitelli, Caserta, Italy}

\affil*[2]{Faculty of Technological and Innovation Sciences, Universitas Mercatorum, Rome, Italy}

\affil[3]{Department of Mathematics and Physics, University of Campania Luigi Vanvitelli, Caserta, Italy}


\abstract{Many conventional statistical and machine learning methods face challenges when applied directly to high dimensional temporal observations. In recent decades, Functional Data Analysis (FDA) has gained widespread popularity as a framework for modeling and analyzing data that are, by their nature, functions in the domain of time. Although supervised classification has been extensively explored in recent decades within the FDA literature, ensemble learning of functional classifiers has only recently emerged as a topic of significant interest. Thus, the latter subject presents unexplored facets and challenges from various statistical perspectives. The focal point of this paper lies in the realm of ensemble learning for functional data and aims to show how different functional data representations can be used to train ensemble members and how base model predictions can be combined through majority voting. The so-called Functional Voting Classifier (FVC) is proposed to demonstrate how different functional representations leading to augmented diversity can increase predictive accuracy. Many real-world datasets from several domains are used to display that the FVC can significantly enhance performance compared to individual models. The framework presented provides a foundation for voting ensembles with functional data and can stimulate a highly encouraging line of research in the FDA context.}






\maketitle


\section{Introduction}
Functional Data Analysis (FDA) refers to situations where sequences of observations over time or space can be suitably represented by functions instead of scalars \cite{Ferraty2003}. The basic premise is to model the underlying function generating the data directly, rather than the sequence of observations, and thus treating the observed functional data as single entities \cite{Ramsay2005}.
FDA has emerged as a collection of statistical methods that can be applied in various analyses, especially in the context of high-dimensional data that is increasingly available in various sectors. The evolution of technology has led to a proliferation of instruments capable of collecting vast volumes of data. This is particularly evident in the domains of biomedical monitoring, Internet of Things (IoT) devices, and finance, which continuously gather data for different purposes. The expansive nature of such data sources, ranging from medical instruments like electrocardiograms (ECG) and electroencephalograms (EEG) to everyday devices like smartphones and IoT devices, highlights the challenge of dealing with this data today.

Over the past few decades, considerable research has explored supervised classification challenges within the FDA literature. However, the intrinsic high-dimensionality existing in functional data leads to the classical issues of the Curse of Dimensionality (COD), which refers to several harmful consequences like sparsity, model choice, explanation challenges, multicollinearity, and distance concentration \cite{bellman1959adaptive}. The complexity of these issues has increased the need for solid statistical methods to analyze and understand such data effectively. 

In the ``non-functional'' framework, dimensionality reduction techniques are commonly used to address the COD. Among these, feature selection methods, such as Recursive Feature Elimination (RFE) and Random Forest Features Importance, can select a subset of the most relevant features for the prediction task. However, the latter techniques can not be directly applicable to sequential data, which requires considering the temporal dependencies and dynamics of the features and might be influenced by correlations. Another option is Principal Component Analysis (PCA), which reduces the dimensionality by projecting the data onto a lower-dimensional subspace spanned by the principal components with the largest singular values. However, PCA may lack interpretability, as the principal components are linear combinations of the original features that may not have a clear meaning or relation to the underlying phenomena. Moreover, using PCA for feature selection can lead to decreased predictive accuracy, as it discards some information from the data that may be relevant to the outcome variable. Therefore, there is a need for more suitable dimensionality reduction methods for sequential data that can balance between interpretability, information preservation, and computational efficiency.
A primary motivation for FDA is that it mitigates the classical issues of high-dimensional data \cite{Ferraty2006} \cite{Cuevas2014}. Indeed, FDA provides dimensionality reduction without restrictive assumptions \cite{Ramsay2005}, making it popular for problems like classifying biomedical time series \cite{maturo2022pooling}  and spatial  functional time series data \cite{medina2014classification}.

In addition to the works that brought FDA to the interest of the international scientific community, such as \citep{Ramsay2002, Ferraty2003, Ramsay2005, Ferraty2011}, there are many additional methodological proposals to extend traditional classification techniques to the context of FDA, e.g. k-Nearest Neighbors (KNN) and logistic regression.
However, in recent decades, tree ensemble methods like Random Forest (RF) \cite{Breiman2004RandomF} and Gradient Boosting (GB) \cite{friedman2001greedy} have become popular for classification and regression problems involving high-dimensional data. Ensemble methods use multiple base learners like Decision Trees (DT) and aggregate their predictions to improve stability and accuracy over a single estimator. For this reason, various approaches have been recently explored also to adapt Classification Trees (CT) and RF to the functional case (see e.g. \cite{Balakrishnan_2006} \cite{Moller2016}).
Yu et al. \citep{Yu_1999} proposed adopting spline trees for functional data, applying time-of-day patterns for customers who place international calls. Nerini et al.
\citep{Nerini_2007} concentrated on the issue of creating a regression tree when the response variable is a probability density function. 
\citep{Fan_2010} dealt with the problem of functional data classification for temporal gene expression data with kernel-induced random forests. Gregorutti et al.
\citep{Gregorutti_2015} focused on the problem of assessing variables' importance. El Haouij et al.
\citep{El_Haouij_2018} offered an extension of the random forest procedure via wavelet basis with an application for driver's stress level.
More recently, Maturo and Verde \citep{maturo2022pooling,maturo2023supervised,maturo2022combining} proposed original extensions of CTs and RF to functional data for improving accuracy and interpretability.

While these demonstrate a growing interest in ensemble methods for functional data, significant gaps exist. More complex ensemble architectures, such as voting ensembles, remain unexplored for functional data.
Key open questions remain about representing functional data, defining criteria for selecting ensemble members, and aggregating predictions from base learners. 
Particularly, in the context of ensemble methods, diversity plays a crucial role.
Diversity refers to the degree of disagreement between the individual learners in an ensemble. In typical machine learning ensembles, training procedures like bagging and boosting intentionally induce diversity among the base learners. Diverse ensembles can achieve lower generalization errors than individual learners by allowing learners to complement each other's weaknesses. Nevertheless, while diversity is crucial, it must be accompanied by accuracy. Ensembles comprising members with significantly low accuracy are unlikely to enhance predictive performance, irrespective of their diversity \cite{good_bad_diversity}.

For functional data, open questions remain regarding how to induce diversity among ensemble members. Data representation is one potential solution, i.e. using different basis function representations (e.g., Fourier, wavelet, B-spline) for different base learners could encourage diversity. The pooling approaches that transform functional data into vectors for classical algorithms, such as functional principal components, may also impact it. However, the impact of diversity in functional data ensembles remains unclear, and how much of it leads to optimal ensemble performance must be studied.
 
Our work aims to address some of these gaps and provide tools for harnessing the power of ensemble techniques in FDA classification problems.
We introduce the so-called Functional Voting Classifier (FVC), i.e. an innovative ensemble architecture designed specifically for functional data. Experimental results on real datasets show the efficacy of ensemble learning in enhancing predictive performance, as measured by accuracy, for functional data when compared to individual models. In addition, as well-known in the statistical learning literature, building an ensemble reduces variance compared to individual models \cite{breiman1996bagging}. By aggregating across multiple diverse base learners, the ensemble helps improve stability and reduce overfitting to noise in the functional training data, highlighting the benefits of ensemble techniques for high-dimensional functional inputs.

The paper is organized as follows. In Section \ref{fda}, we formalize the problem of supervised learning for functional data and review core concepts in FDA. Section \ref{ensemble} surveys ensemble methods and recent advances in applying them to functional inputs. Our proposed ensemble framework for functional data is presented in Section \ref{fvc}. Experimental results on real-world functional classification tasks are given in Section \ref{application}. Section \ref{concl} concludes with a discussion of future research directions.

\section{Functional data analysis} 
\label{fda} 
In many scientific and engineering fields, data is collected in the form of discrete observations or measurements made at particular points in time or space. However, thinking of these discrete points as generated by an underlying continuous function is often desirable. 

Several key benefits of the FDA framework have been highlighted in the literature. FDA is well-suited for inherently functional problems, such as ecological population dynamics \cite{Maturo2018hill, QREI}, climatic variation forecasting \cite{Ramsay2005}, growth curves, and medical research \cite{Ramsay2005}. Moreover, the FDA approach can help assess important additional sources of pattern and variation in data that other techniques may miss \cite{Ferraty2006, Maturo2018nbd}. 
Indeed, FDA provides a functional representation of the phenomenon under study \cite{Ferraty2006}, and thus, crucial information can be captured by modeling the first and second derivatives rather than just the raw data themselves \cite{Ramsay1991}. Another advantage of FDA is that data do not need to be sampled at equally spaced time points, in contrast to time series analysis \cite{Cuevas2014}. Finally, many fundamental notions and theorems of classical statistics can be extended to the infinite-dimensional context of functional data \cite{Cuevas2014, Ramsay2005}.

Let us establish some notation and definitions. Let $\{(t_i, y_i)\}_{i=1}^N$ represent a set of $N$ observations, where $t_i \in 
T$ denotes the feature vectors (observations over time or space), and $y_i \in \mathbb{R}$ is the corresponding response or dependent variable value. The set $T$ is the function's domain, which we assume to be a bounded interval on the real line. The goal is to estimate an underlying function $x(t): T \rightarrow \mathbb{R}$ that generated these observations, defined as $x_i(t) = f(t_i) + \epsilon_i$ where the $\epsilon_i$ represent noise or error terms. 

To convert the discrete points into a functional representation, we need to approximate the function $x(t)$ by some element of a function space. Typically, the latter will be a Hilbert space $H$ consisting of square-integrable functions on $T$ with an inner product defined by

\begin{equation}
\langle x(t),g(t) \rangle = \int_T x(t)g(t) \,dt.
\end{equation}

The norm induced by this inner product is 

\begin{equation}
\|x(t)\| = \sqrt{\langle x(t), x(t) \rangle} = \left(\int_T x^2(t) \,dt \right)^{1/2}.
\end{equation}

This $L^2$ norm quantifies the size of a function. Other norms may also be useful in quantifying properties like smoothness or regularization.

A common approach to representing functional data is to approximate $x(t)$ by a linear combination of basis functions $\{\phi_k\}_{k=1}^K$ given by:

\begin{equation}  \label{fdaaprox}
x(t) \approx \hat{x}(t) = \sum_{k=1}^K c_k \phi_k(t).
\end{equation}

To determine the coefficients from observed data, Least Squares Method, Regularization Techniques, Gradient Descent, Optimization Algorithms, and Cross-Validation Methods can be used. The choice of procedure depends on the specific characteristics of the data and the goals of the analysis. Each method has its advantages and limitations, and the selection often involves a trade-off between accuracy and simplicity.  Polynomials, splines, Fourier series, and wavelets are common choices for basis functions \cite{Ramsay2005}. 
An important factor is deciding how many basis functions $K$ to use. Opting for a small number of basis functions might make the approximation too smooth, introducing a bias. On the contrary, relying on too many basis functions may result in an over-approximation, leading to higher variance \cite{Ramsay2005}. Methods like cross-validation can help determine an appropriate $K$.

B-splines are notably efficacious due to their computational simplicity and flexibility in modeling various functional forms. B-splines, or basis splines, are defined as piecewise polynomial functions of degree \( o \), which are smooth and continuous up to their \( (o-1)^{th} \) derivative. The definition of a spline curve is given by:

\begin{equation}
S(t) = \sum_{k=0}^{K} c_k B_{k,o}(t)
\end{equation}

The construction of a B-spline function, \( B(x) \), involves defining a knot sequence, \( \{t_i\}_{i=1}^{K+o+1} \), where each \( t_i \) is a point in the domain of \( B(x) \) and \( K \) is the number of B-spline basis functions. This sequence partitions the domain into intervals, within which the B-splines are polynomials of degree \( o \). The B-spline basis functions are defined recursively using the Cox-de Boor \cite{de1972calculating} recursion formula:

\begin{equation}
B_{k,0}(x) = 
\begin{cases}
1 & \text{if } t_k \leq x < t_{k+1}, \\
0 & \text{otherwise.}
\end{cases}
\end{equation}

\begin{equation}
B_{k,o}(x) = \frac{x - t_k}{t_{k+o} - t_k} B_{k,o-1}(x) + \frac{t_{k+o+1} - x}{t_{k+o+1} - t_{k+1}} B_{k+1,o-1}(x)
\end{equation}

The functional data is represented as a linear combination of B-spline basis functions:

\begin{equation}
x_i(t) \approx \sum_{k=1}^{K} c_k B_{k,o}(x)
\end{equation}

Here, \( c_i \) are the coefficients that are determined through methods such as least squares fitting, solving:

\begin{equation}
\mathbf{B}^T\mathbf{B}\mathbf{c} = \mathbf{B}^T\mathbf{y}
\end{equation}

Where
\begin{equation}
\mathbf{B} = \begin{bmatrix} B_{1,o}(t_1) & \cdots & B_{K,o}(t_1) \\ \vdots & \ddots & \vdots \\ B_{1,K}(t_K) & \cdots & B_{K,o}(t_K) \end{bmatrix}
\end{equation}

and
\begin{equation}
\mathbf{c} = (c_1, \ldots, c_K)^T
\end{equation}

This representation effectively transforms the problem of analyzing functional data into a finite-dimensional problem, thereby facilitating the application of methodologies discussed in Section \ref{fda}. The primary advantage of using B-splines in FDA is their capacity to provide a flexible yet computationally efficient approximation of complex functional forms. 

B-splines also exhibit the variation diminishing property, ensuring they do not oscillate more than their control polygon and avoiding unwanted ripples. They remain invariant under affine transformations of control points, hence operations like translation and scaling can be applied by transforming the control points alone. Additionally, B-splines are guaranteed to lie within the convex hull of their control points, containing no unwanted loops or oscillations  \cite{de1976splines}. Their basis functions are nonnegative for all parameter values, distinguishing them from other spline types and simplifying optimization algorithms. For open B-splines, the curve is guaranteed to interpolate its endpoint control points, providing control over segment endpoints. Finally, B-splines are at least $C_{o-1}$ continuous, yielding smooth, continuous curves at higher degrees. These attributes make B-splines a convenient modeling tool.
However, selecting the number and placement of knots is a critical aspect that affects the quality of the spline approximation. While too few knots may lead to underfitting, excessively many knots can cause overfitting. Moreover, the choice of the B-spline degree affects the approximation's smoothness, with higher degrees yielding smoother functions but increasing the risk of overfitting.

\subsection{Binary classification in the functional setting} 

The aim of functional classification is to predict an outcome \(Y\) by employing a predictor variable \(X\) taking values in a separable metric space \((E, d)\). In theory, the variable \(Y\) may assume either categorical or numerical values, thereby presenting classification or regression challenges, respectively. However, the present investigation is specifically oriented towards scalar-on-function classification, where \(Y\) is a categorical variable. Thus, the method is intended for functional data of the form \(\{x_i(t), y_i\}\), with a curve \(x_i(t), t \in T\) as the predictor, and \(y_i\) as the response at sample \(i = 1, ..., N\). Let \(Y\) take on the values 0 or 1. A mapping \(f : F \rightarrow \{0, 1\}\), called a \textit{binary classifier}, classifies a new observation \(x_{new}(t)\) from \(X\) by mapping it to its predicted label. This binary case can be extended when \(Y\) has multiple levels. 

Using B-splines, the features matrix is:

\begin{equation} \label{C}
  \textbf{C} =  \begin{bmatrix}  
c_{11} & \dots & c_{1K} \\
\vdots & \ddots & \vdots \\
c_{N1} & \dots & c_{NK} 
\end{bmatrix}
\end{equation}

where \(c_{ik}\) is the coefficient for the \(i\)th curve \(i = 1, ..., N\) relative to the \(k\)th \(k = 1, ..., K\) basis function \(\phi_k(t)\) in the linear combination. 

In predictive modeling, \textbf{C} acts as a dimensionally reduced representation of the original data. This reduction is significant because it transforms the high-dimensional data into a lower-dimensional space without losing critical information. Also, it encapsulates the functional data into a discrete form, which is more manageable. The rows of \textbf{C} represent individual observations, and the columns correspond to the coefficients of the basis functions. This matrix can be directly used in various statistical and machine learning methods for prediction purposes, such as the ones explained in Section \ref{fda}. 
In other words, using basis coefficients provides a significant advantage because it allows the utilization of any classification model by considering the coefficients as features, eliminating the need to define additional distance metrics.

\subsection{Functional Classification Trees} 
Classification trees (CT) are a type of supervised machine learning model used for classification tasks  \cite{Breiman1984trees}. The goal is to predict the value of a categorical target variable based on input features. CTs partition the feature space in a recursive manner, splitting the data into increasingly homogeneous subsets with respect to the target variable. 
Functional Decision Trees (FDTs) extend traditional multivariate decision trees for classification and regression tasks involving functional data inputs. When used for classification tasks, they are called Functional Classification Trees (FCT) \cite{maturo2023supervised}. Rather than operating directly on the raw data, we leverage the coefficients $\{c_{ij}\}$ of the functional data in some fixed basis expansion defined in eq. \ref{fdaaprox}. This approach simplifies the analysis and interpretation of functional data while providing a flexible representation that can capture important features. Using the coefficients as features, the FCT learning process closely mirrors that of classical CT, recursively partitioning the feature space to optimize some splitting criterion. During training, FCTs recursively split the feature space of basis coefficients into rectangular partitions or nodes. The standard top-down greedy algorithm is used, selecting locally optimal splits at each node to maximize some criterion like information gain, reducing node impurity. Each recursive split divides the training data at that node into two child nodes, continuing until stopping criteria are met. The result is a hierarchical binary tree where each internal node represents a split or decision rule, and leaf nodes represent a classification outcome. More specifically, at each node of the tree, the standard decision tree splitting criteria (e.g., information gain, Gini impurity \cite{Breiman1984trees}) can be used to determine the best split, but operating on the basis coefficients $\mathbf{c}_i$ rather than the functional representations directly \cite{maturo2023supervised}. Thus, each split is based on thresholding one of the basis coefficients. The terminal nodes of the tree then define a rectangular partitioning on this coefficients' space $\mathbb{R}^K$, with each rectangle containing collections of coefficient vectors $\mathbf{c}_i$ following decision rules determined by the tree splits. 

FDTs interpretation revolves around the hierarchical splits, which encode nonlinear decision rules on the basis coefficients. Each split isolates a rectangular subregion of the basis coefficient feature space by design, corresponding to a distinct functional data behavior. The path traversed to reach a leaf encodes the conjunction of splits, or functional data attributes, that yield a particular response prediction. An advantage of FDTs is the ability to quantify variable importance, providing insight into which parts of the curve are most relevant for classification. This is straightforward when using a B-spline basis, where each basis function corresponds to a localized region of the function. The standard decision tree variable importance metrics like Gini importance \cite{Breiman1984trees} can be applied to the B-spline coefficients. The coefficients with the highest importance values indicate the regions of the function that provide the most information gain for classification. For example, if higher-order coefficients localized to a specific sub-region are most important, this identifies the locations on the curves that drive the classifications. FDTs provide an essential tool to handle high-dimensional data using multivariate classification trees, but performance and interpretability depend strongly on the basis expansion.

\subsection{Functional k-Nearest Neighbors} 
The k-Nearest Neighbors (KNN) algorithm is used to predict values of responses by looking at the $h$ closest samples in the feature space\footnote{To ensure consistency of the notation and avoid overlap between the symbols used throughout the paper, we refer to $h$ to indicate the hyperparameter of the KNN.}. The Functional k-Nearest Neighbors (FKNN) algorithm extends KNN to the functional framework \cite[see e.g.][]{maturo2022combining}. In this paper, the coefficients of a fixed basis system are used to compute distances between functional data samples. 

Let $\boldsymbol{c}_x$ be the coefficient vector for the new curve \(x_{new}(t)\) and $\boldsymbol{c}_i$ be the coefficient vector for the \(i\)th training curve \(x_i\). The FKNN classifier predicts the class label \(y\) of \(x_{new}(t)\) by finding the $h$ nearest curves in the training set based on the distance between the coefficients vectors: 

\begin{equation}
d(x_{new}, x_i) = \|\boldsymbol{c}_x - \boldsymbol{c}_i\|
\end{equation}

Consider a labeled dataset $T$ consisting of $N$ tuples $(t_i, y_i)$, where $t_i \in \mathbb{R}^T$ is a $T$-dimensional feature vector and $y_i \in \{0, 1, \ldots, Z\}$ is the corresponding label. The aim is to classify a new unlabeled instance $x_{new}(t)$.

First, the coefficient matrix \textbf{C} defined in eq. \ref{C} is extracted. The feature space is now the $\mathbb{R}^K$ space spanned by the rows of \textbf{C}, and $S$ is a set of labeled instances transformed similarly. Given a new, unlabeled instance $x_{new}(t)$ with $\boldsymbol{c} \in \mathbb{R}^K$, the FKNN algorithm first identifies the $h$ points in $S$ that are nearest to $\boldsymbol{c}$. The distance between $\boldsymbol{c}$ and $\boldsymbol{c}_i$ is typically calculated using a distance metric. The default distance metric used in KNN is the Minkowski distance, which is defined as:

\begin{equation}
d(x, x_i) = \left(\sum_{j=1}^{s} \|c_{j} - c_{ij}\|^v\right)^{1/v}
\end{equation}

where $c_j$ and $c_{ij}$ are the $j$-th components of $c$ and $c_i$ respectively, and $v$ is a parameter (default $v=2$ which represents Euclidean distance).

The $h$ nearest neighbors of $x_{new}(t)$ are the $h$ curves $\{x_{i1}(t), x_{i2}(t), \ldots, x_{ih}(t)\}$ for which $d(x, x_{ij})$ is the smallest.

The FKNN classification rule then assigns $x_{new}(t)$ a label $\hat{y}$ as follows:

\begin{equation}
\hat{y} = \arg\max_{c} \sum_{i=1}^{k} I(y_{ij} = u) \cdot w(i)
\end{equation}

where $u$ is a variable taking values in the set $\{0, 1, \ldots, Z\}$, where $Z$ represents the number of categories of $Y$. \(I(\hat{y}_{ij} = u)\) is the indicator function defined as \(I(\hat{y}_{ij} = u) = 1\) if \(\hat{y}_{ij} = u\) and 0 otherwise, and \(w(i)\) is a weighting function assigning a weight to the \(i\)-th neighbor. In the case of KNN, the weights can be uniform (\(w(i) = 1\) for all \(i\)), inversely proportional to distance (\(w(i) = 1/d(x, x_{ij})\)), or a user-defined function.

Therefore, FKNN states that $x_{new}(t)$ is assigned to the class $u$ for which the weighted sum of the indicators $I(y_{ij} = u)$ is maximized. Hence, $x_{new}(t)$ is assigned to the most common class among its $h$ nearest neighbors.

\subsection{Functional Random Forest}

Functional Random Forest (FRF) \cite{maturo2022combining} extends the concepts of random forest to functional data classification and regression problems.  As in FDTs, the raw functional observations ${x_i(t)}$ are projected onto a fixed basis ${\phi_k(t)}$ to obtain coefficient features ${c_{ik}}$. FRF then leverages an ensemble of decision trees, each trained on bootstrap samples of the coefficients $\boldsymbol{C} = {c_{ik}}$.

The FRF training process follows the standard Random Forest algorithm \cite{Breiman2004RandomF}, with a few key differences in handling functional inputs:

\begin{enumerate}
\item For each tree in the ensemble $b = 1, \dots, B$:
\begin{enumerate}
\item Draw a bootstrap sample of size $N$ from the training set
\item Grow a full FDT tree $T_b$ on the bootstrap sample, recursively partitioning the coefficient feature space. However, rather than evaluate all possible splits at each node, randomly sample $m$ try of the $K$ total coefficients as split candidates.
\end{enumerate}
\item Make predictions by aggregating the predictions from the $B$ trees. For classification, this involves majority voting across the tree predictions.
\end{enumerate}

By training each FDT on a slightly different bootstrap sample and evaluating splits on random subsets of coefficients, FRF injects additional diversity into the ensemble, reducing variance and improving generalizability. The key parameters controlling the FRF complexity are the number of trees $B$, the number of randomly sampled split candidates $m$, and the depth to grow each tree before stopping. Increasing $B$ reduces the variance but increases computational expense. The value of $m$ introduces randomness in the tree building, with smaller $m$ leading to greater diversity. The tree depth controls the maximum interactions considered, avoiding overfitting. FRF variable importance metrics can also be calculated on the coefficients to quantify which basis functions are most relevant for classification. For B-spline bases, this identifies key regions of the functional data driving the ensemble predictions. Compared to FDTs, FRF predictions exhibit lower variance by aggregating across multiple trees \cite{maturo2022pooling}. By randomly sampling a subset of coefficient features to consider at each split, rather than evaluating splits on all coefficients, FRF trees inject additional randomness into the tree building process. This means the decision rules in any single FRF tree are less deterministic and more variable than in a corresponding FDT grown on the same data \cite{Breiman2004RandomF}. Hence, an individual FRF tree will be less interpretable, as the splits are based on random samples of coefficients rather than optimal splits across all coefficients, as in FDTs.

\subsection{Functional Gradient Boosting}

Gradient boosting machines (GBMs) are an ensemble technique that produces a prediction model in the form of an expansion of weak learners, usually decision trees \cite{friedman2001greedy}. Functional gradient boosting extends GBMs to leverage functional data by operating on the basis coefficients as features. 

Given training data $\{(\mathbf{c}_i, y_i)\}_{i=1}^N$ consisting of basis coefficient vectors $\mathbf{c}_i \in \mathbb{R}^K$ and binary class labels $y_i \in \{0,1\}$, the functional GBM learns an additive expansion:

\begin{equation}
F(\mathbf{c}) = \sum_{b=1}^B f_b(\mathbf{c}) 
\end{equation}

where the $f_b(\mathbf{c})$ are the weak learners fit sequentially and $B$ is the number of boosting iterations. Typically $f_b(\mathbf{c})$ are decision trees or regressions trees, giving a sum-of-trees model.

Functional Gradient Boosting (FGB) is derived from the perspective of numerical optimization in function space \cite{friedman2001greedy}. The goal is to find the function $F(\mathbf{c})$ that minimizes the expected value of some specified loss function $\mathcal{L}(y, F(\mathbf{c}))$:

\begin{equation}
F^* = \arg min_{F} \mathbb{E}_{y, \mathbf{c}} [ \mathcal{L}(y, F(\mathbf{c})) ]
\end{equation}

For binary classification, common loss functions are logistic loss for probability estimation or exponential loss for classification. Functional gradient descent iteratively adds components $f_b(\mathbf{c})$ that point in the negative gradient direction, sequentially improving the loss. Initially $F_0(\mathbf{c}) = 0$, then at iteration $b$:

\begin{align}
f_b(\mathbf{c}) &= \argmin_{f} \sum_{i=1}^N [- \nabla_{\mathbf{c}} \mathcal{L}(y_i, F_{b-1}(\mathbf{c}_i)) \cdot f(\mathbf{c}_i) ] \\
F_b(\mathbf{c}) &= F_{b-1}(\mathbf{c}) + \nu f_b(\mathbf{c})
\end{align}

Where $\nu$ is a learning rate. Each $f_b(\mathbf{c})$ is fit to the negative gradient or pseudo-residuals using the current model $F_{b-1}(\mathbf{c})$. This process greedily minimizes the loss by adding increments in the gradient direction. In practice, each weak learner is fit on a random subsample of the training data for computational and statistical efficiency. Shrinkage is also commonly used, scaling each $f_b$ by a small factor $\nu < 1$ to reduce overfitting. The functional gradient descent perspective provides a principled approach to fitting an additive expansion predictive model to basis coefficients of functional data. The components $f_b$ can capture nonlinear effects and interactions between basis coefficients to build a stronger learner. Tree-based methods for the components provide interpretability, with each tree split indicating an interaction between basis coefficients relevant to the gradient loss reduction. The relative influence of basis coefficients can also be quantified via the total number of splits or gain metrics.

\section{Ensemble learning and the key role of diversity} 
\label{ensemble}
Ensemble learning refers to combining multiple machine learning models to obtain improved predictive performance. The core concept is that an ensemble of diverse and independently trained learners can outperform any constituent model. By aggregating predictions across a set of models, the individual learners' strengths can be leveraged while their weaknesses are overcome. Two popular ensemble techniques are bagging and boosting. Bagging involves training models on random subsets of the data and then combining predictions by voting or averaging \cite{breiman1996bagging}. By sampling the data, bagging exploits variation across learners to reduce variance. Boosting sequentially trains models on reweighted versions of the data to focus on previously misclassified instances. Through iterative learning, boosting reduces bias. Both approaches capitalize on model diversity to enhance predictions. 

Diversity refers to the degree of disagreement between the individual learners that make up an ensemble. Greater diversity leads to a more robust ensemble that can provide improved predictions compared to individual models \cite{kuncheva2003measures}. There are two main types of diversity: data diversity and model diversity. Data diversity involves training individual learners on different subsets or representations of the training data. Common approaches for inducing data diversity include bagging, where models are trained on bootstrap samples of the data, and partitioning the input features so that different models see different subsets of features. Instead, model diversity involves using different learning algorithms and model hyperparameters to allow the learners to capture distinct aspects of the problem. 

Diversity can be quantified in several ways. Pairwise diversity measures like Q-statistics \cite{kuncheva2003measures} and disagreement measure the proportion of instances on which two models make different predictions. Higher values indicate greater diversity between the pair. Global diversity metrics like entropy \cite{kuncheva2003measures} and Kohavi-Wolpert variance \cite{Kohavi1996BiasPV} look at diversity across all ensemble members and measure the degree to which models make errors on different instances. 
Jaccard distance \cite{kuncheva2004distance} is another useful pairwise diversity measure defined as: 

\begin{equation} 
\label{dj}
d_J = 1 - \frac{|A $\cap$ B|}{|A $\cup$ B|}
\end{equation}

Equation \ref{dj} represents the dissimilarity between two model predictions sets $A$ and $B$ based on their intersection over union. Values range from 0 (identical) to 1 (completely different). More diverse ensembles exhibit greater entropy and variance. 

Several theoretical analyses demonstrated the link between diversity and ensemble performance \cite{good_bad_diversity}. Error-ambiguity decomposition shows that ensemble error is lower when constituent models make uncorrelated errors on test instances. The bias-variance trade-off motivates using more diverse but higher variance models, as their errors can cancel out in the ensemble. Diversity must be balanced with individual model accuracy: overly diverse but weak learners will not improve performance. 

In the FDA approach, diversity may arise because different base approximations may provide unique functional perspectives on the data, naturally leading models to capture divergent characteristics and increasing disagreement.

\subsection{Voting classifiers in ensemble learning}

Voting refers to building an ensemble of multiple classifiers, aggregating their outputs to get the final prediction. The aggregation function can be a majority vote (hard voting) or the average of the probabilities for each class (soft voting) \cite{geron2017hands}.

Diversity in a voting ensemble represents the degree of difference among the predictions of individual models contributing to the ensemble. This variety is key to robust decision-making, allowing the ensemble to tap into a broader range of perspectives and capture more information. An increased ensemble diversity is associated with improving the majority vote \citep{kuncheva2003measures}. This is due to the \textit{wisdom of the crowd} effect, i.e. when models make uncorrelated errors, these mistakes are likely to be canceled out during the ensemble's majority voting process.

In order to create a diverse ensemble, there are multiple strategies. In a voting ensemble, one can augment diversity by training models on different versions of the datasets or by using different models.
One method is to train models on different versions of the datasets.  This approach encourages the base models to learn different aspects or features of the data, making their decision boundaries vary, hence promoting diversity \citep{breiman1996bagging}. This strategy can involve various techniques like bootstrapping the samples, perturbing the input features, or varying the data preprocessing steps. Another effective technique to introduce diversity into a voting ensemble is through model heterogeneity, i.e., employing different machine learning algorithms as base models. Each algorithm has its strengths and weaknesses. Thus, this diversity helps capture a wide range of data patterns. Moreover, this approach can help alleviate the negative impact of the no-free-lunch theorem, which states that no single algorithm works best for every problem \citep{wolpert1997no}.

One must carefully design the ensemble to ensure that the increased diversity does not lead to detrimental effects such as increased complexity without significant performance improvements. Data diversification refers to using different versions or representations of the training data to train the individual models in an ensemble. The choice between data diversification and model diversification can also depend on the specific problem at hand, the computational resources available, and the interpretability requirements of the model. While diversification through data or models has been widely used, the optimal degree and method of diversity in voting ensembles is still a topic of active research.  A hybrid approach might involve using data augmentation to artificially increase the effective size of the dataset and combining it with model diversification techniques to ensure a rich set of models contributing to the ensemble. 

This paper presents the idea that training ensemble members on different functional representations of the original dataset leads to enhanced predictive performance in classification problems. Specifically, our method, presented in Section \ref{fvc} diversifies the data by training each ensemble member on a different functional basis approximation of the original dataset. Training ensemble members on different functional basis approximations is a form of data diversification that allows the models to focus on different aspects of the data during training.

\section{Supervised classification via ensembles of different functional representations: the Functional Voting Classifier (FVC)}
\label{fvc}

Let us define a training set $\{(x_i,y_i)\}_{i=1}^N$ where $\boldsymbol{x}_i = (x_{i1},...,x_{iP})$ is a $P$-dimensional vector of predictor variables for observation $i$, and $\boldsymbol{y}_i \in \{1,\dots, Z\}$ is the categorical response variable with $Z$ classes. 

To obtain multiple functional representations, we first approximate each vector $\mathbf{x}_i$ as a function $x_i(t)$ using B-spline basis expansions of orders $o=1,\dots, O$:

\begin{equation} 
x_i^{(o)}(t) = \sum_{j=1}^{K^{(o)}} c_{ij}^{(o)} B_j^{(o)}(t)
\end{equation}

where $K^{(o)}$ is the number of B-spline bases used for order $o$. For each order $o$, we employ $Q$-fold cross-validation on the training set to select the optimal $K^{(o)}$.  For each candidate $K^{(o)}$, we compute the cross-validation score by training on $Q-1$ folds and validating on the held-out fold, repeated over all folds. The $K^{(o)}$ that minimizes the average cross-validation error is selected. Using too few bases may lead to underfitting, while too many may overfit. This data-driven approach aims to balance bias and variance. 

Let $\mathbf{C}^{(o)} \in \mathbb{R}^{N \times K^{(o)}}$ be the B-spline coefficient matrix for order $o$, obtained by stacking the coefficient vectors $\mathbf{c}_i^{(o)}$ as rows. 

For each order $o$, the coefficients $\mathbf{C}^{(o)}$ are used to train a supervised learning model $f^{(o)}: \mathbb{R}^{K^{(o)}} \rightarrow \{1,\dots,Z\}$, where any classification algorithm can be implemented (e.g., FKNN, FCT, and FRF). Using different algorithms for each model can further improve diversity. The result is an ensemble of $O$ models $\{f^{(1)},\dots,f^{(O)}\}$, where each $f^{(o)}$ is trained on the functional representations from B-spline order $o$.  For a new test function, we obtain representations $\mathbf{c}^{(o)}$ under each optimal B-spline basis. These are fed into the trained models $f^{(o)}$ to obtain predicted class labels $\hat{y}^{(o)} = f^{(o)}(\mathbf{c}^{(o)})$.

Let $\mathbf{F} \in \mathbb{R}^{M \times O}$ be the prediction matrix containing the predictions from the $O$ models on the $M$ test examples. The element $\mathbf{F}_{mo} = \hat{y}_n^{(o)}$ is the predicted class label from model $o$ on test input $x_m(t)$:

\begin{equation}
\mathbf{F} = \begin{bmatrix}
\hat{y}_1^{(1)} & \hat{y}_1^{(2)} & \dots & \hat{y}_1^{(O)} \\
\hat{y}_2^{(1)} & \hat{y}_2^{(2)} & \dots & \hat{y}_2^{(O)} \\
\vdots & \vdots & \ddots & \vdots \\
\hat{y}_M^{(1)} & \hat{y}_M^{(2)} & \dots & \hat{y}_M^{(O)}
\end{bmatrix}
\end{equation}

The majority vote criteria is used to select the final prediction:

\begin{equation}
\hat{y}_m = \arg\max_{u \in {1,\dots,Z}} \sum_{o=1}^O \mathbb{I}(\mathbf{F}_{mo} = u)
\end{equation}

Where $\mathbb{I}(\mathbf{F}_{mo} = u)$ is an $M$-dimensional indicator vector that has a 1 for test examples where model $o$ predicted class $u$, and 0 otherwise. Summing these binary indicator vectors over $o$ and taking the argmax over $u$ gives the majority vote label for each test example $m$. The final prediction $\hat{y}_n$ is the majority vote across the ensemble members. By combining models trained on different functional representations, this ensemble methodology leverages diversity to improve robustness. The cross-validation procedure provides a data-driven approach to balance underfitting and overfitting within each learner. The hard voting exploits disagreement between models to obtain a more robust prediction. The overall ensemble classifier is expected to outperform individual constituents.

\section{Application} 
\label{application}
The proposed FVC is tested on different datasets. The training data consisted of discrete multivariate observations $\{(x_i,y_i)\}_{i=1}^N$ as described in Section \ref{fvc}. To generate multiple functional representations, each $x_i$ is approximated using B-spline bases of orders $O=\{3,5,7,9,11\}$. For each order $o \in O$, 10-fold cross-validation is used on the training data to select the optimal number of basis functions $K^{(o)}$ that minimized the cross-validation error. 
In order to achieve a good trade-off between computational efficiency and modeling accuracy, a grid search is performed over $K^{(o)}$ using the following criteria:

\begin{equation}
K^{(o)} \in \left\{\frac{N}{20}, \frac{2N}{20}, \ldots, \frac{N}{2} \right\}
\end{equation}

Where $N$ is the number of training observations. The process grids over values of $K^{(o)}$ from $\frac{N}{20}$ to $\frac{N}{2}$ in steps of $\frac{N}{20}$. The lower bound of $\frac{N}{20}$ aims to provide enough bases to sufficiently represent the functional data, while the upper bound of $\frac{N}{2}$ aims to limit computational complexity while achieving a minimum of 50\% reduction in dimensionality compared to the original training data. The interval spacing of $\frac{N}{20}$ provides a reasonable resolution to select an optimal $K^{(o)}$ that balances representation capacity with efficiency. This grid search allows traversing a range of model complexities to find the best trade-off between computation time and modeling accuracy for a given dataset. The optimal $K^{(o)}$ is chosen as the one that minimizes cross-validation error across this grid. 

Four types of classification models are trained on each functional representation: FKNN, FCT, FGB, and FRF. For a given model type (e.g., FKNN), five models are trained on the coefficient matrices $\mathbf{C}^{(o)}$ from each B-spline basis expansion. For a new test observation $x_m(t)$, the coefficient vectors $\mathbf{c}^{(o)}$ are obtained under each of the B-spline bases. The predictions are made by passing each $\mathbf{c}^{(o)}$ into its associated trained model, yielding $\hat{y}^{(o)}$ for $o \in O$. Finally, the predictions are aggregated across models using FVC.

By training an ensemble of classifiers on different B-spline function representations, and combining their predictions through hard voting, we aim to improve robustness and accuracy compared to any individual model.
\subsection{Datasets}
Table \ref{table-datasets} displays the details of the proposed datasets. The datasets are specifically selected to cover a wide range of characteristics and problem domains to test the methods rigorously. In total ten datasets are used, spanning fields including medicine, engineering, science, and finance, taken from the UCR time series archive \cite{tsc};
however, the datasets also cover additional functional data types - \textit{DistalPhalanxOutlineCorrect} (DPOC) consists of data from images, while \textit{MedicalImages} involves classifying histogram shapes. 

The datasets exhibit substantial variability in the key properties of time series classification tasks. The series lengths range from 80 to 720 time points, encapsulating both short and longer sequences. The number of rows for training and testing also vary widely, with some datasets like \textit{ElectricDevices} having thousands of instances and others like Coffee having only dozens. Class balance is also captured, with the ratio of majority to minority class spanning approximately 5:1 to 1:1 across datasets.
For datasets that originally contained more than two classes, the classes have been binarized by setting the non-zero classes as 1, allowing the use of binary classification evaluation metrics.
The training and test sets have no overlap in terms of duplicate instances, and techniques were selected using cross-validation on only the training data to evaluate real-world generalization ability on fully unseen data.

\begin{table}[ht]
\centering
\begin{tabular}{lrrrrr}
\toprule
\textbf{Dataset}                      & \textbf{Length} & \textbf{Train size} & \textbf{Test size} & \textbf{Classes} & \textbf{Balance} \\ 
\midrule
ChlorineConcentration                 & 166           & 467                 & 3840               & 2                & [0.81 0.19]            \\
Coffee                                & 286           & 28                  & 28                 & 2                & [0.5 0.5]             \\
Computers                             & 720           & 250                 & 250                & 2                & [0.5 0.5]             \\
DPOC           & 80            & 600                 & 276                & 2                & [0.37 0.63]           \\
ECG200                                & 96            & 100                 & 100                & 2                & [0.31 0.69]           \\
ECG5000                               & 140           & 500                 & 4500               & 2                & [0.65 0.35]           \\
ElectricDevices                       & 96            & 8926                & 7711               & 2                & [0.75 0.25]           \\
FordA                                 & 500           & 3601                & 1320               & 2                & [0.51 0.49]           \\
GunPoint                              & 150           & 50                  & 150                & 2                & [0.48 0.52]           \\
MedicalImages                         & 99            & 381                 & 760                & 2                & [0.47 0.53]           \\
\bottomrule
\end{tabular}
\caption{\centering Datasets \cite{tsc}}
\label{table-datasets}
\end{table}

\subsection{Different functional representation leading to increased diversity} 
\label{diversity}

Let us denote the B-spline functional representations as $B3, B5, B7, B9,$ and $B11$ respectively, for bases of increasing order $O=\{3,5,7,9,11\}$. For each basis expansion, we train a separate classification model $f^{(o)}(x_i(t))$. These models are the ensemble members for the FVC explained in Section \ref{fvc}.

\begin{figure}
    \centering
    \includegraphics[width=0.7\linewidth]{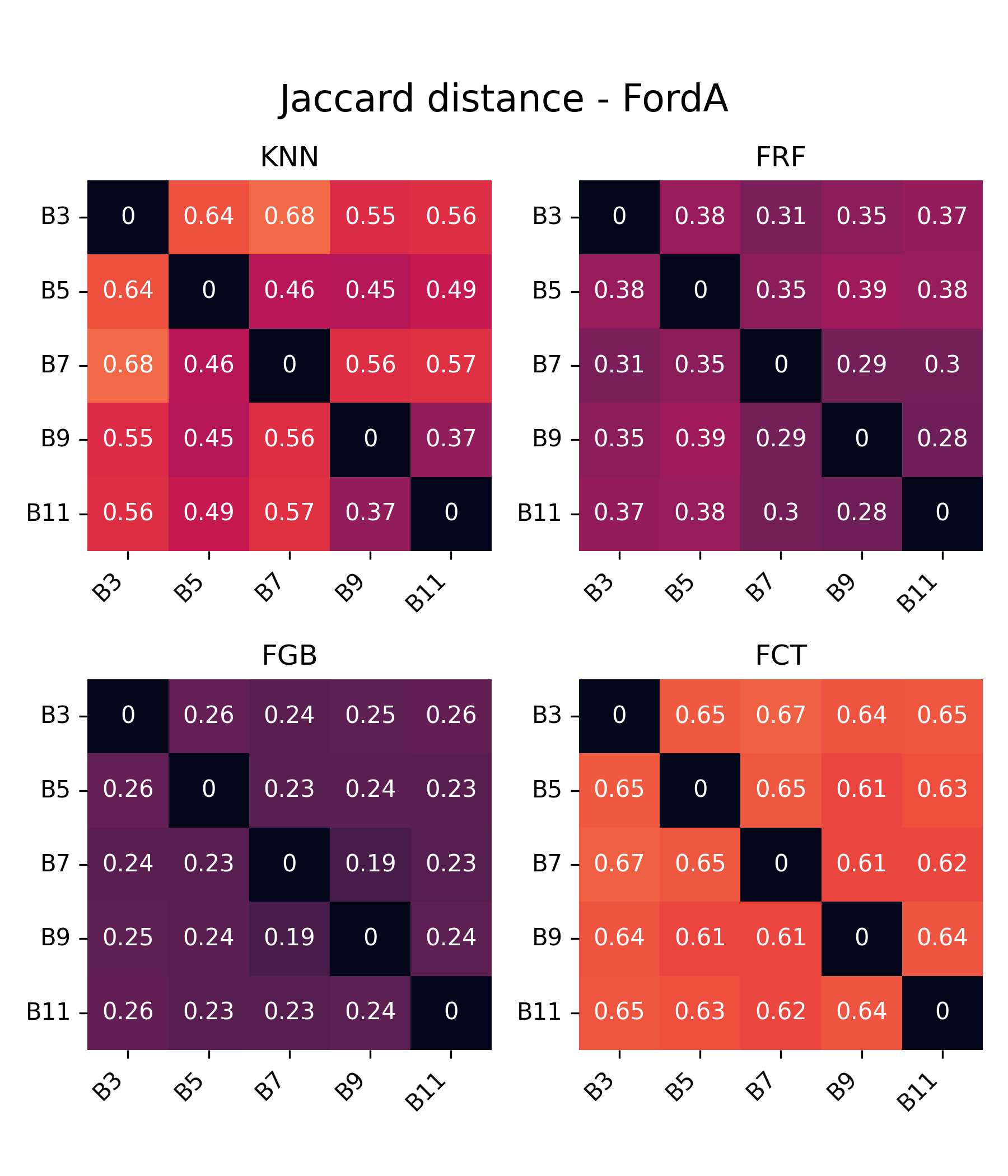}
    \caption{Diversity on the FordA dataset.}
    \label{fig:forda_div}
\end{figure}

\begin{figure}
    \centering
    \includegraphics[width=0.7\linewidth]{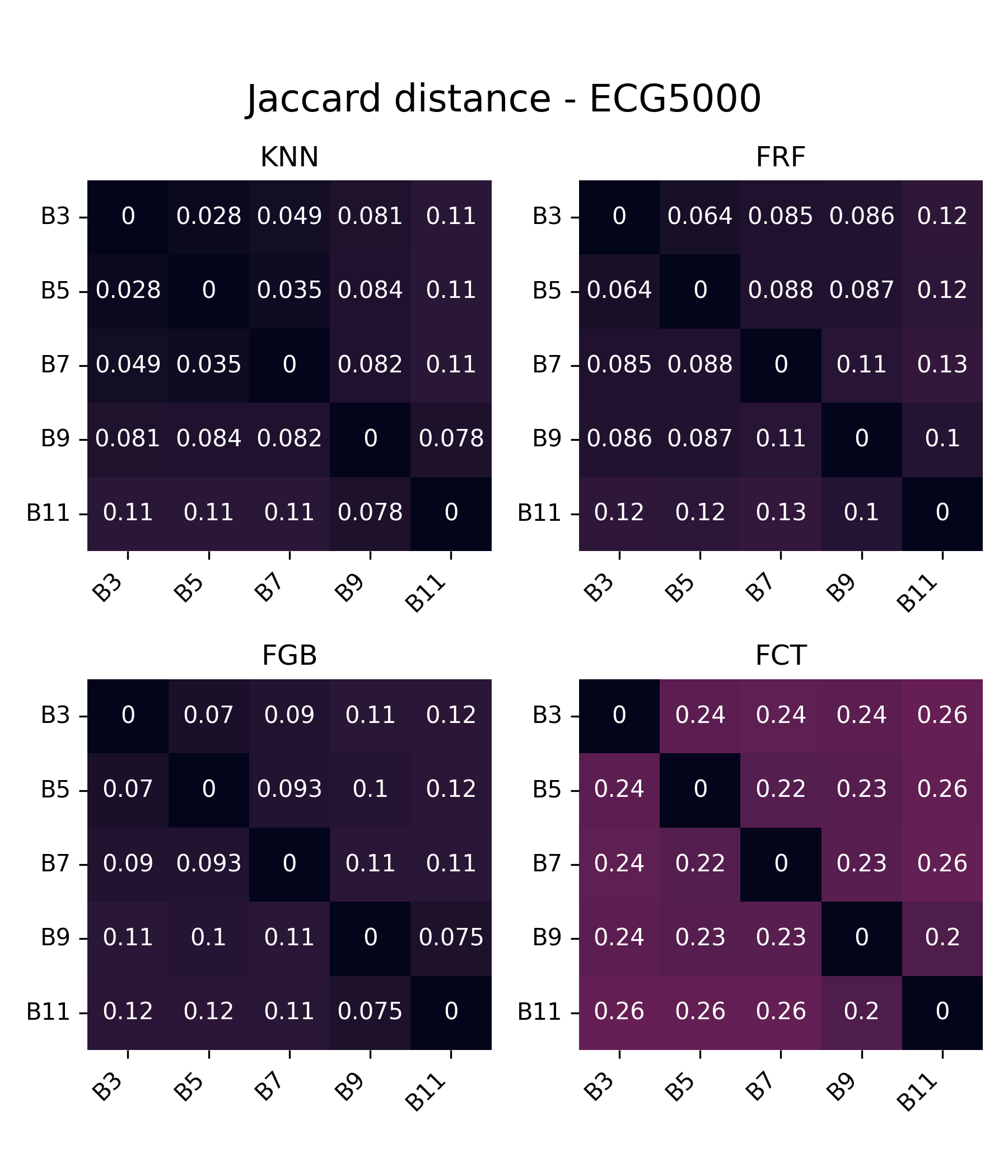}
    \caption{Diversity on the ECG5000 dataset.}
    \label{fig:ecg_div}
\end{figure}

Fig. \ref{fig:forda_div} and \ref{fig:ecg_div} show diversity between ensemble members' predictions measured as Jaccard distance. The results demonstrate that utilizing different B-spline function representations for each base classifier increases diversity within the FVC ensemble, as shown in the Jaccard distance heatmaps between binary predictions from models trained on the various basis expansions. On the \textit{FordA} dataset (Fig \ref{fig:forda_div}), we see a substantial divergence between the predictions of the constituent models. For instance, the KNN classifiers trained on \textit{B3} and \textit{B5} bases have a Jaccard distance of 0.64, indicating that the predictions differ on over 2/3 of test cases, evidencing significant diversity. The same trend holds across all classifier types, with average distances ranging from 0.26 (FGB) to 0.65 (FCT). The increased diversity arises because the different B-spline bases provide unique functional perspectives on the same underlying data. The various basis orders extract distinct patterns and features when approximating each observation as a function. Hence, models trained on these alternate representations will naturally capture different characteristics and lead to greater disagreement. This idea was a key motivation behind the FVC methodology. In contrast, the classifiers show less diversity on the \textit{ECG5000} dataset (Fig. \ref{fig:ecg_div}). The average Jaccard distances range from 0.08 (FGB) to 0.19 (FCT), indicating greater consensus between predictions. However, some degree of complementarity remains, with no two models making identical predictions across all cases. Notably, the diversity pattern holds consistently for each dataset and classifier algorithm. 

Comparing the heatmaps in Fig.\ref{fig:forda_div} and \ref{fig:ecg_div}, \textit{FordA }shows substantially higher distances than \textit{ECG5000}. This finding suggests that the diversity arising from the functional representations depends intrinsically on the properties of the data itself. In cases like \textit{ECG5000} where the bases extract similar features, the ensemble members will naturally agree more often. Nevertheless, the central trend remains - utilizing different basis expansions increases heterogeneity across constituent models.

Other studies with additional datasets (Appendix \ref{appendix_heatmaps}) reveal the same phenomenon. Despite some variability in the extent of diversity in all cases, multiple functional representations provide useful complementary perspectives that improve ensemble performance. Approximating each observation through B-splines of varying order induces diversity between base classifiers in the FVC framework. This process complements the cross-validation approach for optimizing model complexity within each representation. The combination allows for capturing distinct patterns in the data, leading to more robust predictions when aggregated through voting. The consistency of this effect further validates the efficacy of the proposed methodology.

\subsection{Functional Voting Classifier's Results}
In this section, FVC's performance is evaluated by benchmarking its performance against the individual constituent models on a suite of binary classification problems presented in Table \ref{table-datasets}. Tables \ref{table-gb}, \ref{table-knn}, \ref{table-rf}, \ref{table-ct} report the classification accuracy for four model types - FGB, FCT, FKNN, and FRF. The results demonstrate the benefits of ensembling diverse functional representations within the FVC framework. Across all model types and datasets, the FVC ensemble achieves the highest accuracy in most cases. Out of the 40 total configurations, the FVC attains the best performance in 28 settings. Combining predictions from models built on different B-spline basis representations generally improves robustness and accuracy. 

Looking at the FRF results in Table \ref{table-rf}, FVC achieves the highest accuracy on 4 out of 10 datasets (\textit{Coffee, ECG5000, FordA, MedicalImages}), with gains up to 4.5\%. However, for some datasets like \textit{Computers and GunPoint}, the FVC accuracy is comparable to or lower than the best individual FRF model, indicating that the FVC does not always improve over the best base classifier. Notably, the relative gains from ensembling are higher when the base learners display greater diversity. As discussed in Section \ref{diversity}, the different functional perspectives induced by varying the B-spline order lead to more complementary predictions. FVC shows substantial improvements on datasets like \textit{FordA} where the constituent models exhibit high disagreement. This effect is most prominent in the FCT results shown in Table \ref{table-ct}, where FVC produces the highest accuracy on eight datasets, with gains up to 6-7\% compared to the individual model, showing the benefits of ensembling for some problems. 

Examining the FGB results in Table \ref{table-gb}, the FVC again achieves the top accuracy on six datasets. On \textit{Coffee} there is perfect agreement between the ensemble members, and thus the voting ensemble retains this accuracy. There is a substantial improvement of over 3\% in \textit{FordA} and \textit{GunPoint}. Given the average Jaccard distance of 0.26 between FGB constituents, their disagreement helps compensate for errors. Since the bases capture distinct characteristics, cases misclassified by one model may be corrected by others. In contrast, gains are marginal for datasets like \textit{ECG5000} where the base models show high consensus. With average Jaccard distances under 0.09, the constituents provide limited complementary information. Thus, simple voting provides little benefit over the best-performing member. Still, FVC matches the top accuracy, demonstrating that the ensemble methodology does not degrade performance in low-diversity settings. 

The FKNN results (Table \ref{table-knn}) follow similar patterns, confirming the relationship between base model heterogeneity and ensemble performance. For \textit{ECG500} and\textit{ ECG200}, the ensemble accuracy fails to improve over the best base result since the agreement is very high. However, for \textit{FordA}, \textit{GunPoint}, and \textit{DPOC} with higher diversity, the FVC substantially improves over individual learners by 2-3\% absolute.

\begin{table}
\centering
\begin{tabular}{lrrrrrr}
\toprule
Dataset & B3 & B5 & B7 & B9 & B11 & FVC \\
\midrule
ChlorineConcentration & 0.8060 & \textbf{0.8099} & 0.7995 & 0.8016 & 0.7896 & 0.8044 \\
Coffee & \textbf{1.0000} & 0.9643 & \textbf{1.0000} & \textbf{1.0000} & 0.9286 & \textbf{1.0000} \\
Computers & 0.5720 & 0.6120 & \textbf{0.6480} & 0.6320 & 0.6360 & 0.6200 \\
DPOC & 0.7681 & 0.7536 & 0.7717 & 0.7572 & \textbf{0.7754} & \textbf{0.7754} \\
 ECG200 & \textbf{0.8800} & 0.8600 & 0.8000 & 0.7800 & 0.7700 & 0.8100 \\
 ECG5000 & 0.9529 & 0.9500 & 0.9458 & 0.9478 & 0.9442 & \textbf{0.9553} \\
ElectricDevices & 0.9027 & 0.8970 & 0.9043 & 0.8806 & 0.8774 & \textbf{0.9117} \\
FordA & 0.8023 & 0.8205 & 0.8182 & 0.8061 & 0.8091 & \textbf{0.8652} \\
GunPoint & \textbf{0.9467} & 0.8600 & 0.9133 & 0.9000 & 0.7333 & 0.9200 \\
MedicalImages & 0.8026 & 0.7868 & 0.8211 & 0.8013 & 0.8013 & \textbf{0.8276} \\
\bottomrule
\end{tabular}
\caption{\centering \label{table-rf}Accuracy for Functional Random Forest. BO: B-spline representation of order O}
\end{table}

\begin{table}
\centering
\begin{tabular}{lrrrrrr}
\toprule
Dataset & B3 & B5 & B7 & B9 & B11 & FVC \\
\midrule
ChlorineConcentration & 0.8125 & 0.8232 & 0.8021 & 0.8044 & 0.8203 & \textbf{0.8260} \\
Coffee & \textbf{0.5357} & \textbf{0.5357} & \textbf{0.5357} & \textbf{0.5357} & \textbf{0.5357} & \textbf{0.5357} \\
Computers & 0.6120 & 0.5920 & \textbf{0.6160} & 0.6080 & \textbf{0.6160} & 0.6080 \\
DPOC & \textbf{0.7971} & 0.7790 & 0.7754 & 0.7500 & 0.7464 & 0.7790 \\
 ECG200 & \textbf{0.8400} & 0.7900 & \textbf{0.8400} & 0.7000 & 0.7500 & 0.8000 \\
 ECG5000 & 0.9522 & 0.9529 & 0.9473 & 0.9471 & 0.9420 & \textbf{0.9560} \\
ElectricDevices & 0.9231 & 0.9169 & 0.9007 & 0.8774 & 0.8793 & \textbf{0.9271} \\
FordA & 0.8598 & 0.8500 & 0.8500 & 0.8712 & 0.8508 & \textbf{0.8811} \\
GunPoint & 0.8133 & 0.8267 & 0.8267 & 0.8267 & 0.6533 & \textbf{0.8667} \\
MedicalImages & 0.7737 & 0.8026 & 0.7895 & 0.7276 & 0.7579 & \textbf{0.8092} \\
\bottomrule
\end{tabular}
\caption{\centering \label{table-gb}Accuracy for Functional Gradient Boosting. BO: B-spline representation of order O}
\end{table}

\begin{table}
\centering
\begin{tabular}{lrrrrrr}
\toprule
Dataset & B3 & B5 & B7 & B9 & B11 & FVC \\
\midrule
ChlorineConcentration & 0.7729 & 0.7451 & 0.7201 & 0.7224 & 0.7367 & \textbf{0.7982} \\
Coffee & 0.9286 & \textbf{1.0000} & \textbf{1.0000} & \textbf{1.0000} & 0.9286 & \textbf{1.0000} \\
Computers & 0.5480 & 0.5720 & 0.5000 & 0.5600 & \textbf{0.5920} & 0.5640 \\
DPOC & 0.7319 & 0.6848 & 0.7101 & 0.7283 & 0.7138 & \textbf{0.7645} \\
 ECG200 & \textbf{0.8300} & 0.8100 & 0.7800 & 0.7400 & 0.7100 & 0.8000 \\
 ECG5000 & 0.9220 & 0.9138 & 0.9229 & 0.9129 & 0.9102 & \textbf{0.9378} \\
ElectricDevices & 0.8701 & 0.8641 & 0.8680 & 0.8492 & 0.8468 & \textbf{0.9083} \\
FordA & 0.6023 & 0.6152 & 0.6295 & 0.6659 & 0.6530 & \textbf{0.7121} \\
GunPoint & 0.7933 & 0.7667 & 0.8000 & 0.7467 & 0.8000 & \textbf{0.8600} \\
MedicalImages & 0.7237 & 0.6724 & 0.7066 & 0.6961 & 0.6816 & \textbf{0.7658} \\
\bottomrule
\end{tabular}
\caption{\centering \label{table-ct}Accuracy for Functional Classification Tree. BO: B-spline representation of order O}
\end{table}

\begin{table}
\centering
\begin{tabular}{lrrrrrr}
\toprule
Dataset & B3 & B5 & B7 & B9 & B11 & FVC \\
\midrule
ChlorineConcentration & 0.7380 & 0.7513 & 0.7492 & 0.7547 & 0.7292 & \textbf{0.7589} \\
Coffee & 0.9643 & 0.9643 & \textbf{1.0000} & \textbf{1.0000} & 0.8929 & \textbf{1.0000} \\
Computers & 0.5360 & 0.5840 & 0.6000 & 0.6080 & 0.5960 & \textbf{0.6160} \\
DPOC & 0.7138 & 0.7101 & 0.7210 & 0.7174 & 0.7138 & \textbf{0.7319} \\
 ECG200 & 0.8800 & \textbf{0.8700} & 0.8200 & 0.7900 & 0.7900 & 0.8400 \\
 ECG5000 & \textbf{0.9540} & 0.9529 & 0.9489 & 0.9460 & 0.9367 & 0.9531 \\
ElectricDevices & 0.8809 & 0.8794 & 0.8627 & 0.8444 & 0.8407 & \textbf{0.8861} \\
FordA & 0.7144 & 0.7152 & 0.7030 & 0.7348 & 0.7402 & \textbf{0.7750} \\
GunPoint & 0.7667 & 0.8133 & 0.8400 & 0.8533 & 0.8400 & \textbf{0.8600} \\
MedicalImages & \textbf{0.7579} & 0.7289 & 0.7237 & 0.6882 & 0.7013 & 0.7539 \\
\bottomrule
\end{tabular}
\caption{\centering \label{table-knn}Accuracy for Functional k-Nearest Neighbors. BO: B-spline representation of order O}
\end{table}



\section{Discussion and conclusions}
\label{concl}

This research introduces the FVC, i.e. an ensemble methodology tailored for supervised classification problems involving functional data inputs. Rather than relying on raw multivariate observations, the proposed FVC suggests representing data as functions approximated under different bases in order to exploit the information power of a diverse ensemble. This study shows that by training constituent models on the different functional representations and aggregating their predictions through voting, the FVC improves accuracy and robustness compared to individual learners. The experimental results demonstrate several key benefits of the proposed approach. The diversity in datasets and tasks further enhances the rigor and generalizability of the benchmark evaluation. Performance across these distinct problem settings quantifies flexibility and applicability to multifaceted functional data. The FVC attained state-of-the-art accuracy across a suite of real-world binary classification tasks, consistently outperforming the best individual model in most benchmark experiments. For certain datasets such as \textit{FordA} and \textit{GunPoint}, the FVC provided absolute gains in accuracy up to 6-7\% over the top-performing constituent. The latter result highlights the efficacy of synergizing predictions from diverse models trained on distinct functional perspectives of the data.

The degree of improvement also depends on the base model complexity. With simple KNN models, diversity plays an even more significant role in compensating for individual weaknesses. The cross-validation procedure helps mitigate the latter issue, allowing small gains even with complex FGB models. Overall, more significant heterogeneity appears more beneficial for less flexible base learners. The FVC demonstrates state-of-the-art accuracy by synthesizing diverse predictions from base models trained on different functional representations. It provides an elegant way to learn complementary patterns within the data through cross-validated B-spline approximations. This ensemble methodology is widely applicable across classifier algorithms and time series datasets. The benchmark results validate FVC's ability to leverage diversity for improved robustness and accuracy compared to individual constituent models.

Importantly, this research also revealed insights into the effects of diversity within functional ensembles. FVC's accuracy improvements were most prominent in cases where the base learners exhibited high diversity in their predictions, as quantified by Jaccard distance. This finding demonstrates empirically that increased heterogeneity between functional representations leads to greater ensemble performance, aligning with diversity theories in machine learning. Improvements for multiple base model types, including FGB, FRF, FCT, and FKNN evidenced the flexibility of the FVC framework. The simple voting aggregation improves robustness by exploiting uncorrelated errors between the diverse ensemble members. FVC retains a close connection to the original data distribution since diversity is introduced through functional representations rather than direct sampling. The cross-validation procedure further avoids overfitting within each basis expansion. The framework is also modular and compatible with any classification algorithm applied to the spline coefficients, enabling FVC for different tasks and domains. FVC delivers an accurate ensemble methodology for functional classification, offering dimensionality reduction and implicit regularization. 

A potential drawback is reduced interpretability compared to a single model due to the multiple representations and voting aggregation. However, model-agnostic variable importance measures could help extract insights. In addition, there is an increased computational expense of training multiple models compared to a single classifier. However, this cost can be justified by the potential accuracy and robustness gains demonstrated experimentally. While demonstrating promising results on several datasets, FVC has some limitations that could guide future work. The choice of basis type is fixed rather than learned adaptively. Meta-learning techniques could potentially optimize the ensemble architecture and functional representations in a data-driven manner. Opportunities also exist to incorporate greater diversity through heterogeneous models, stacked ensembles, and more sophisticated aggregation methods than voting. Additional research on associating basis functions with domain knowledge could further enhance model interpretability. Theoretical analysis of diversity and generalization error could provide valuable insights into the underpinnings of the model's efficacy. Nevertheless, this research presented an important first realization of voting-based methodology for functional data classification. The FVC framework establishes a foundation for advancing functional data analysis through diverse ensemble learning.
 
\section*{Competing Interests}
The authors have no relevant financial or non-financial interests to disclose.

\section*{Ethical approval}
This article does not contain any studies with human participants performed by any of the authors.


\newpage

\section{Appendix}

\subsection{Diversity plots}
\label{appendix_heatmaps}

\begin{figure}[!htb]
    \centering
    \includegraphics[width=0.6\linewidth]{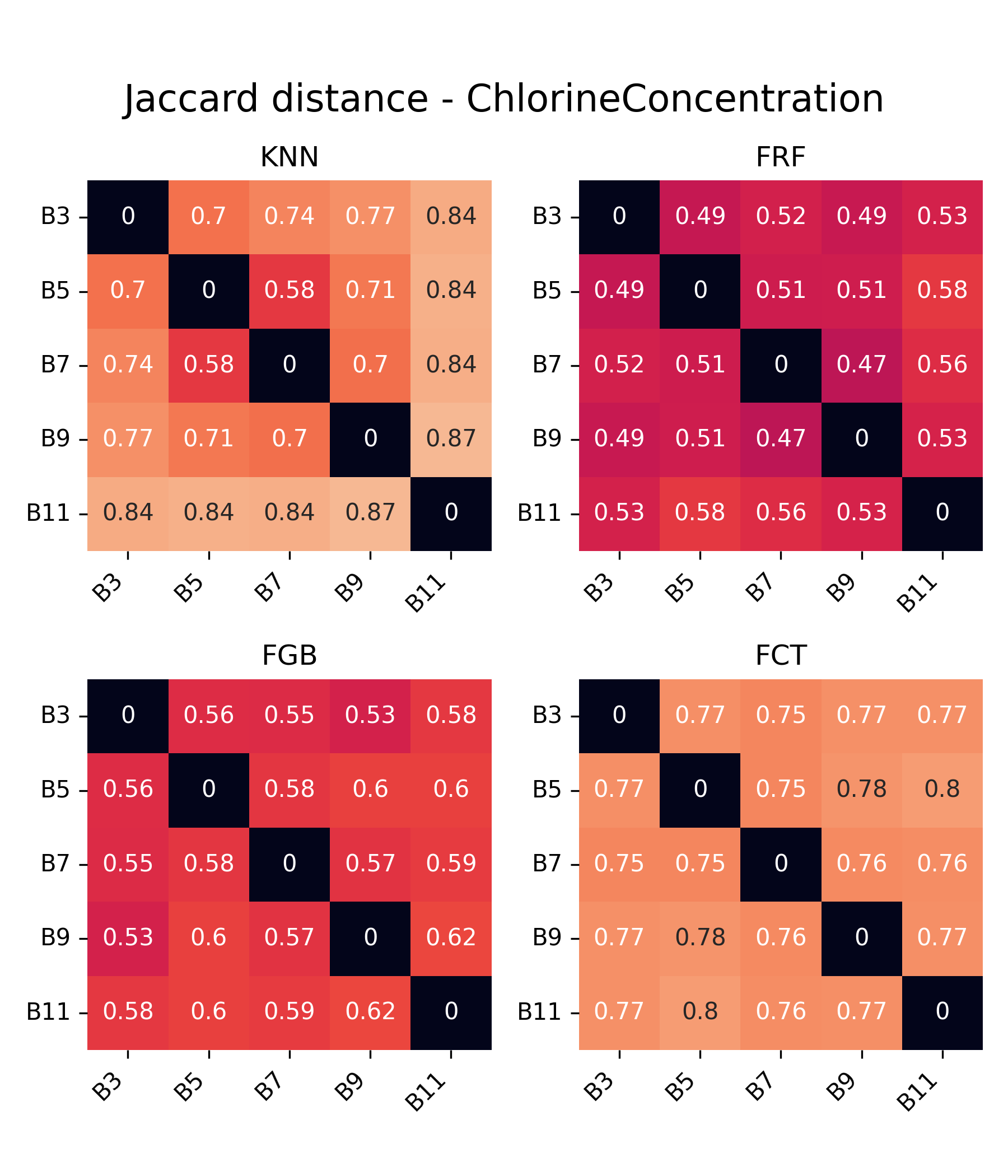}
    \caption{Diversity on the ChlorineConcentration dataset.}
\end{figure}

\begin{figure}[!htb]
    \centering
    \includegraphics[width=0.6\linewidth]{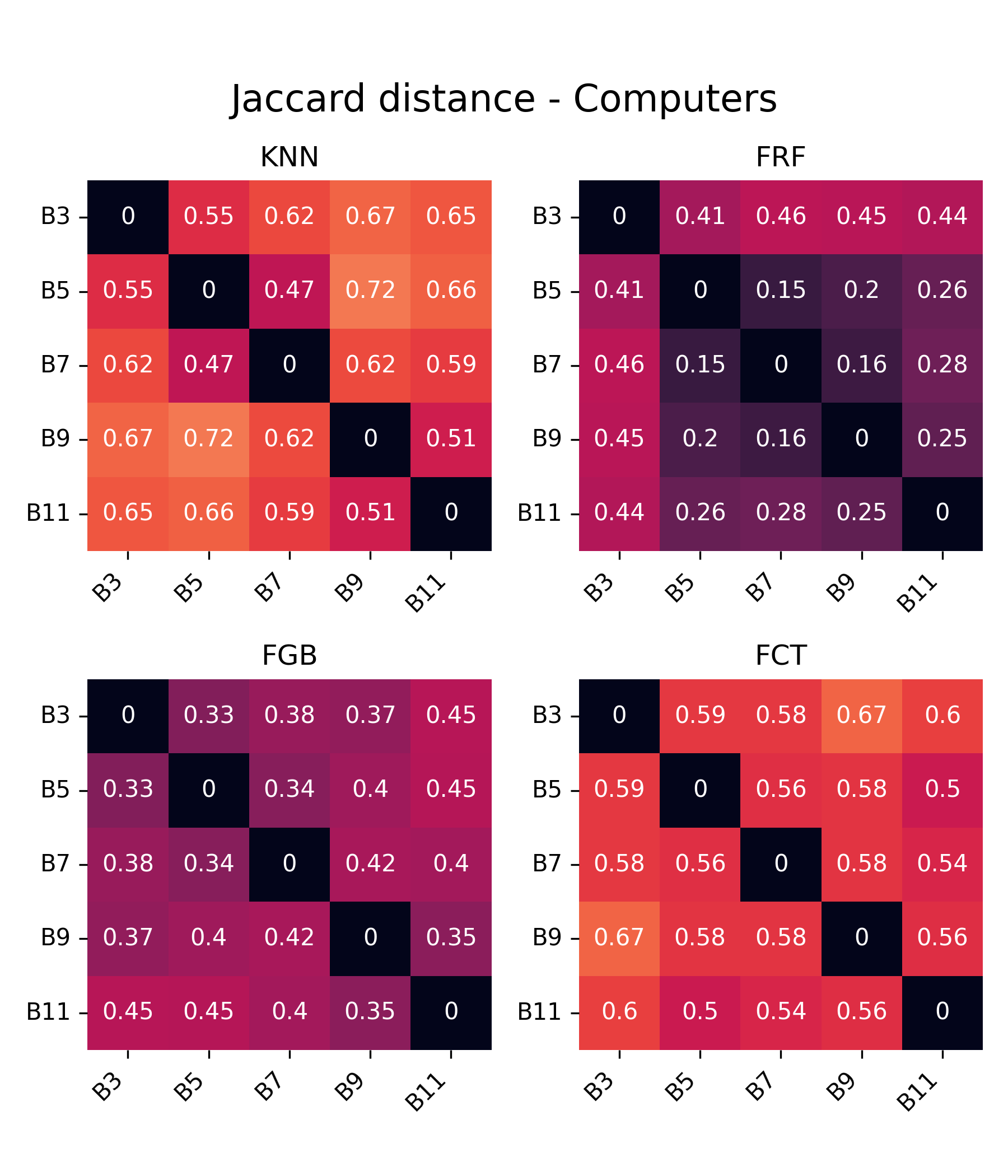}
    \caption{Diversity on the Computers dataset.}
\end{figure}

\begin{figure}[!htb]
    \centering
    \includegraphics[width=0.6\linewidth]{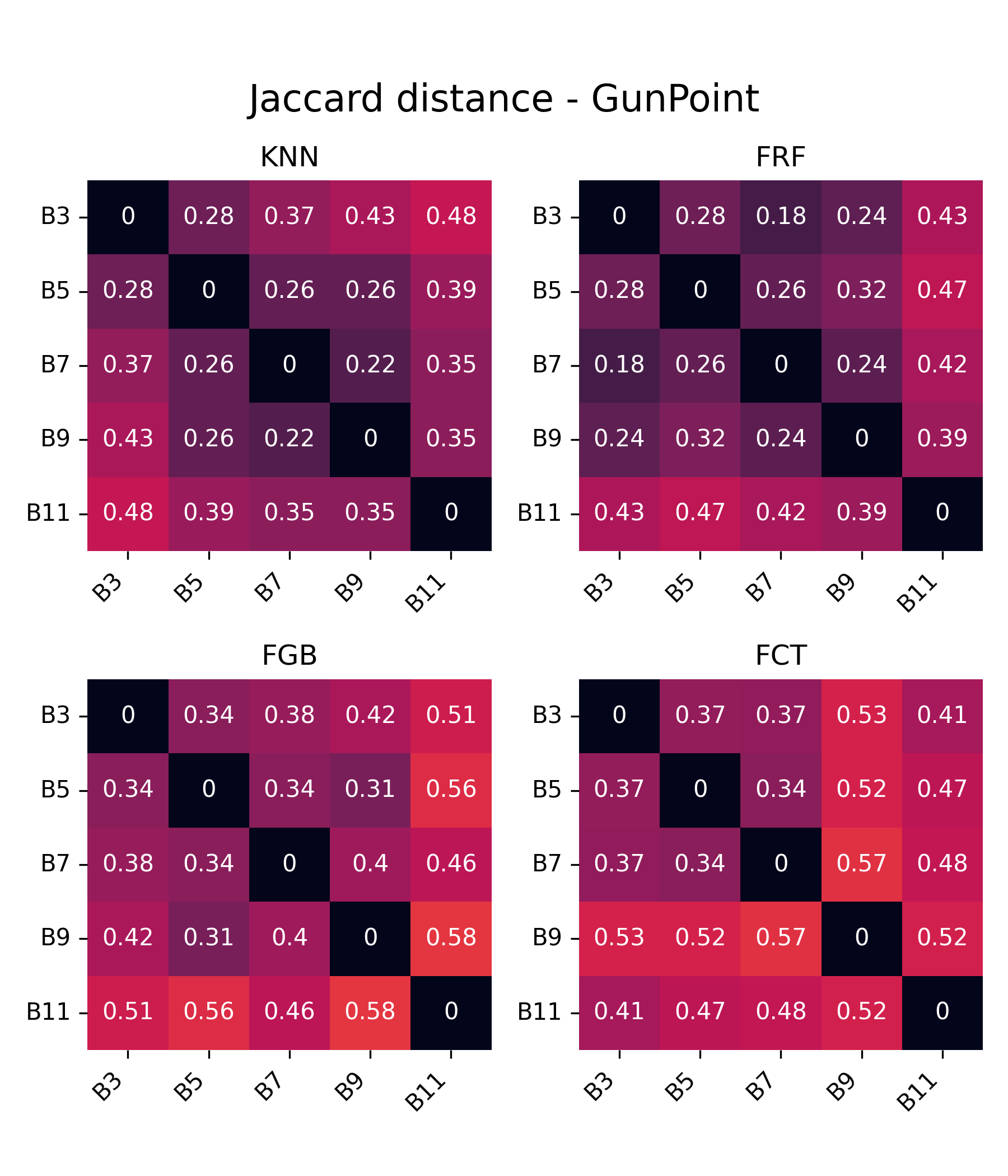}
    \caption{Diversity on the GunPoint dataset.}
\end{figure}

\begin{figure}[!htb]
    \centering
    \includegraphics[width=0.6\linewidth]{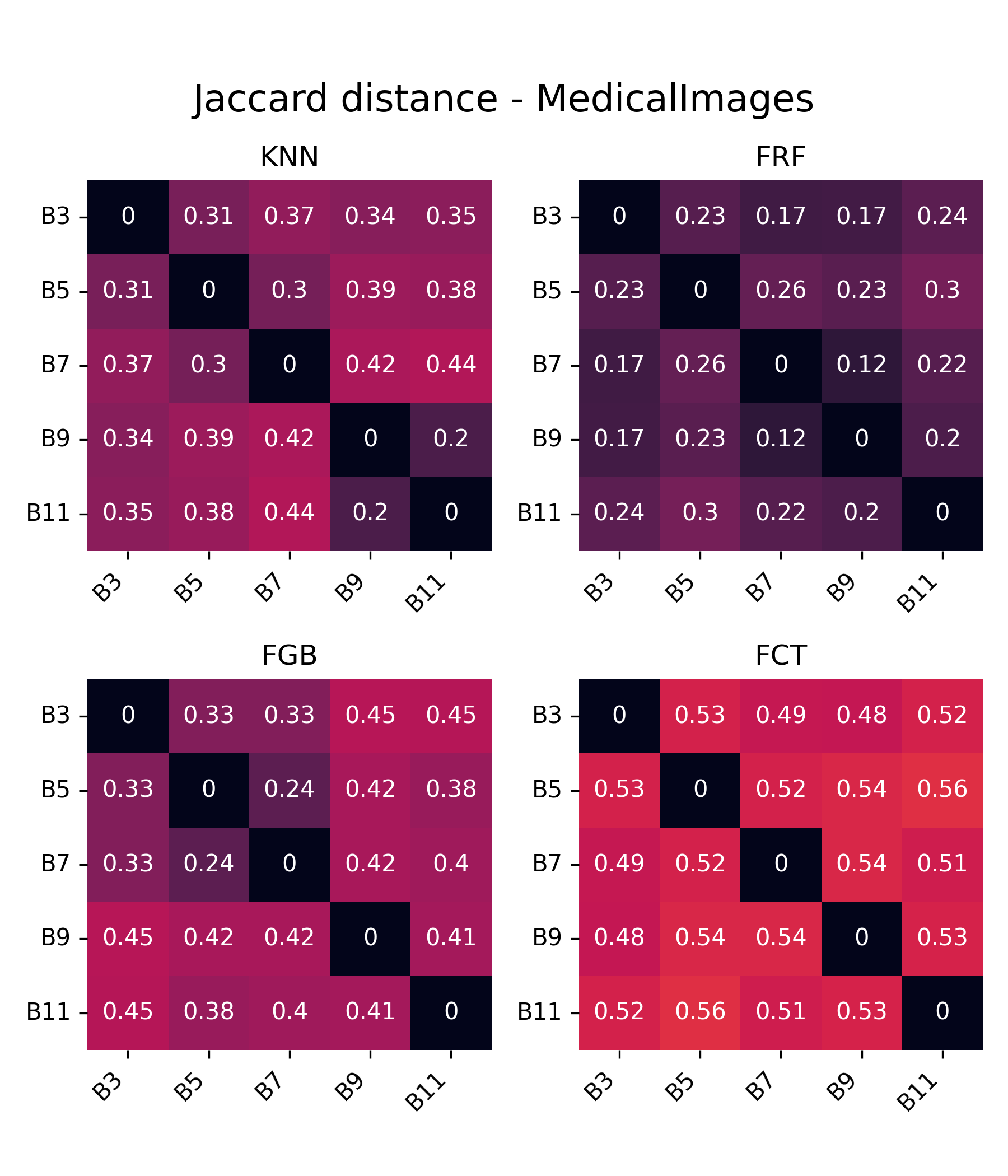}
    \caption{Diversity on the MedicalImages dataset.}
\end{figure}

\begin{figure}[!htb]
    \centering
    \includegraphics[width=0.6\linewidth]{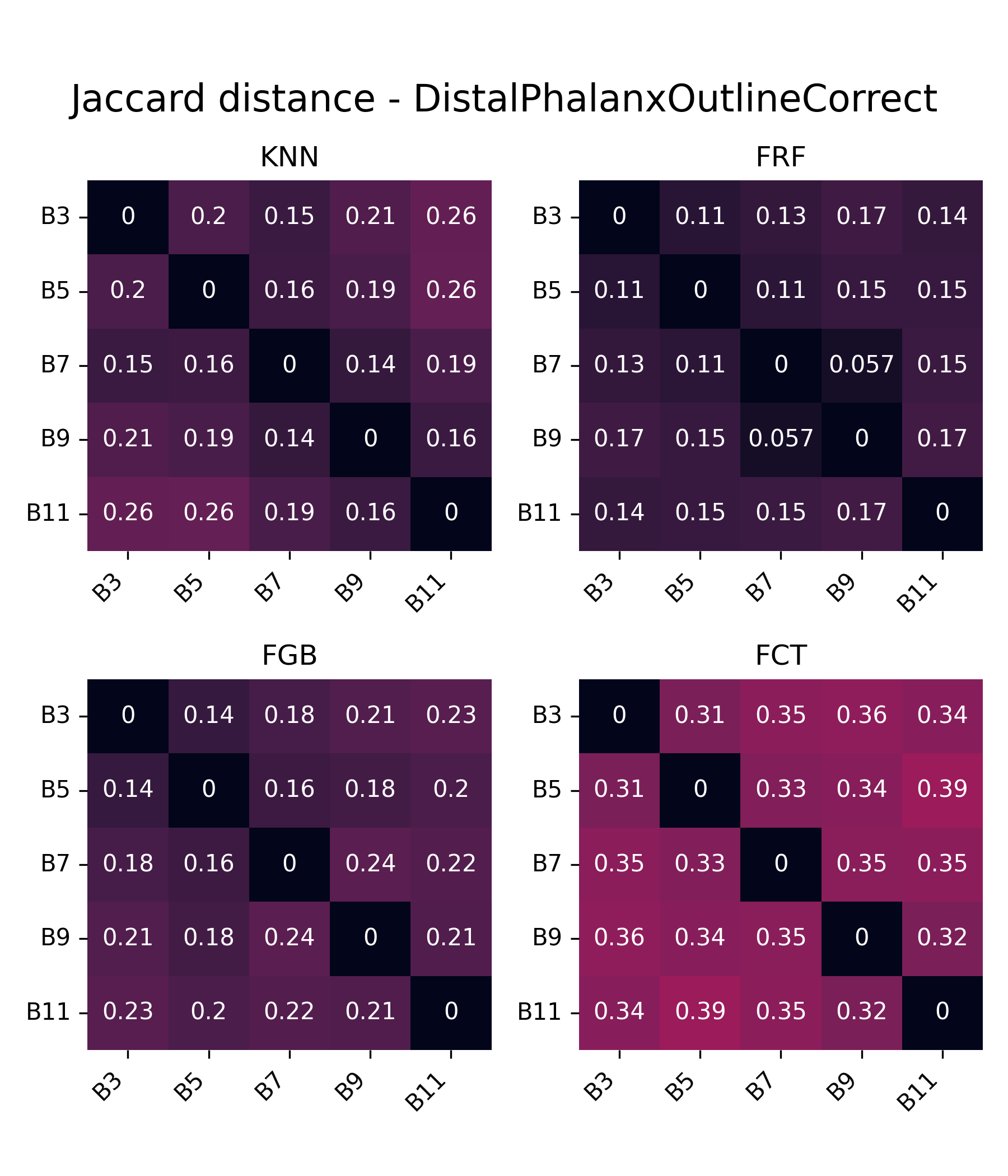}
    \caption{Diversity on the DistalPhalanxOutlineCorrect dataset.}
\end{figure}

\begin{figure}[!htb]
    \centering
    \includegraphics[width=0.6\linewidth]{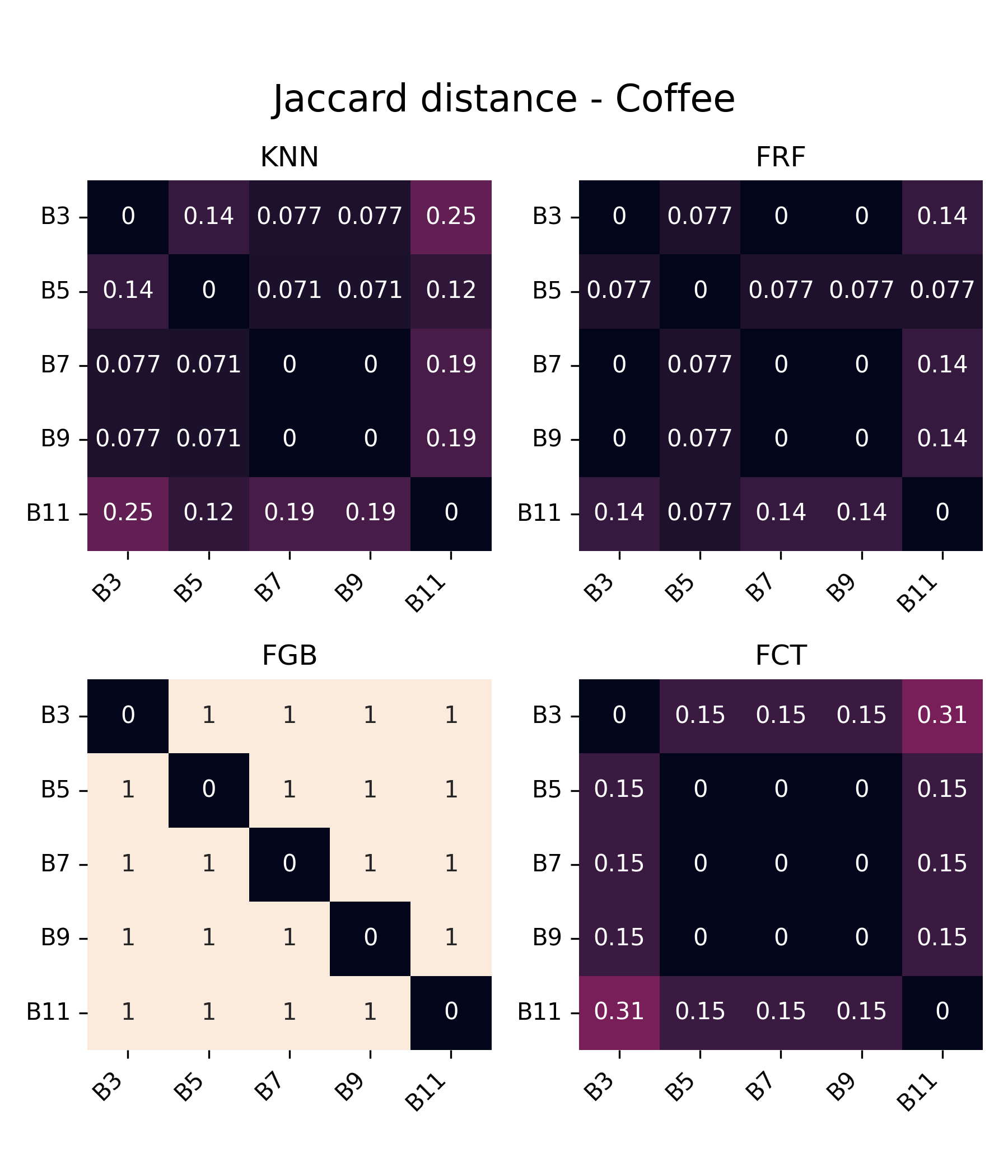}
    \caption{Diversity on the Coffee dataset.}
\end{figure}

\begin{figure}[!htb]
    \centering
    \includegraphics[width=0.6\linewidth]{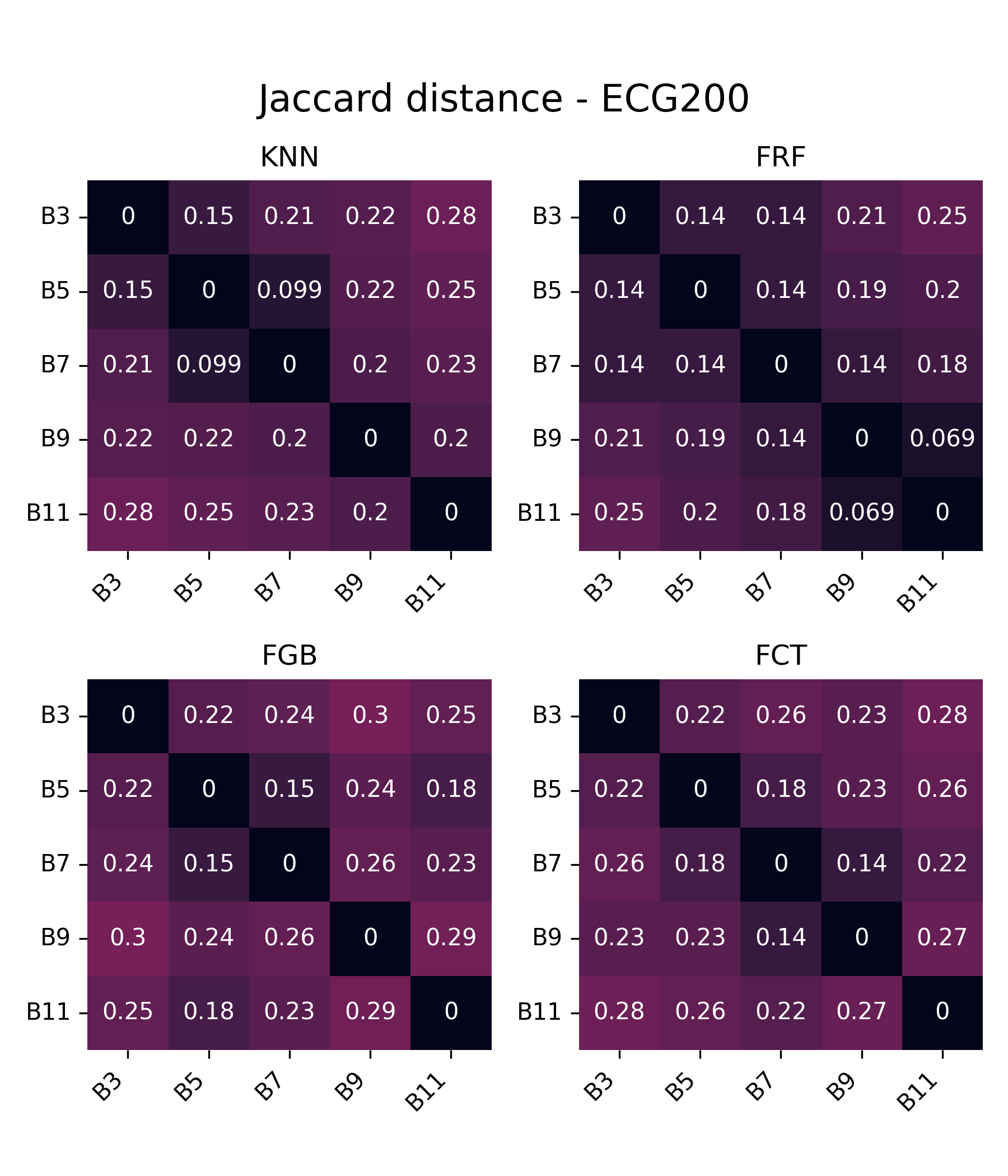}
    \caption{Diversity on the GECG200 dataset.}
\end{figure}

\begin{figure}[!htb]
    \centering
    \includegraphics[width=0.6\linewidth]{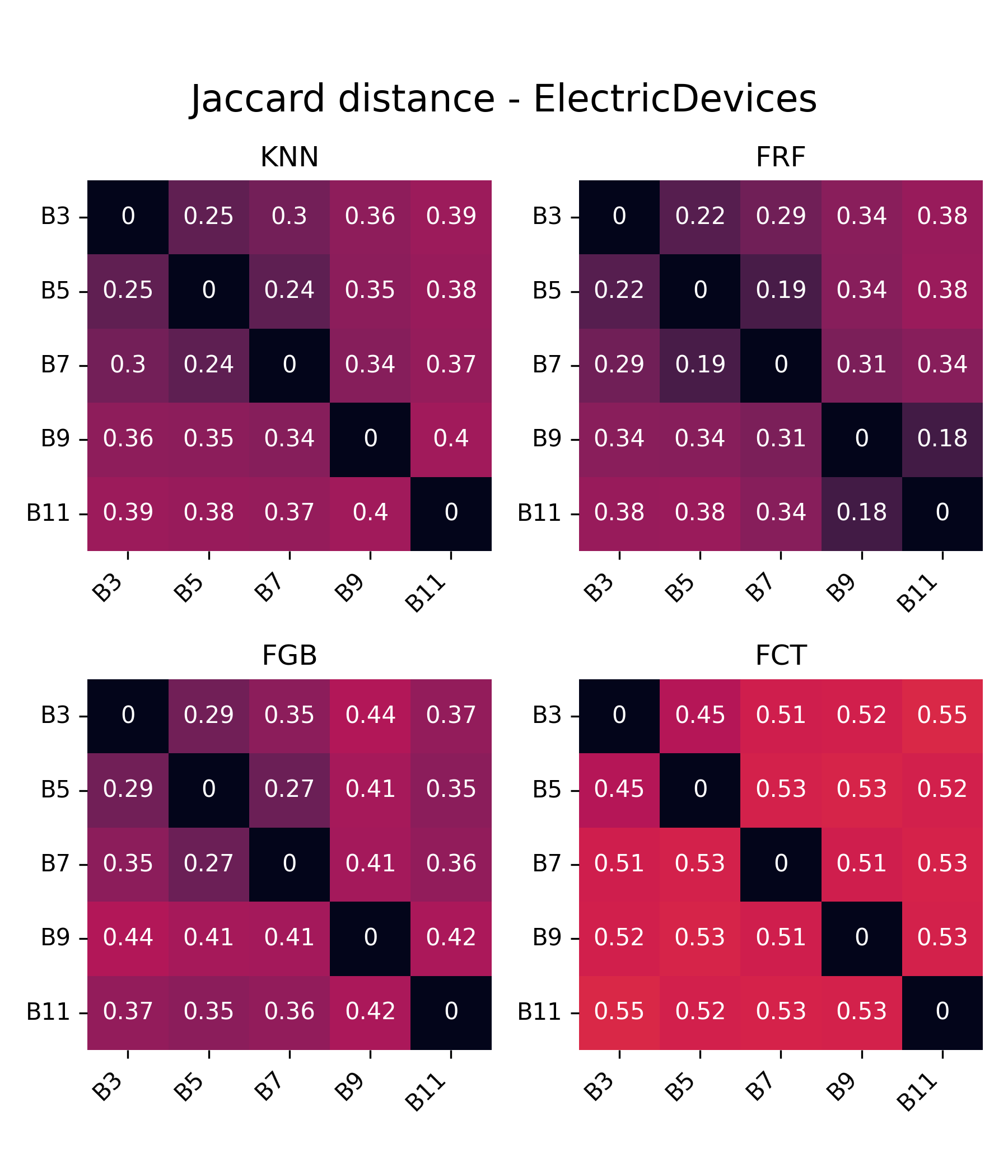}
    \caption{Diversity on the ElectricDevices dataset.}
\end{figure}

\clearpage

\subsection{Functional representations}
\begin{figure}[!htb]
    \centering
    \includegraphics[width=1\linewidth]{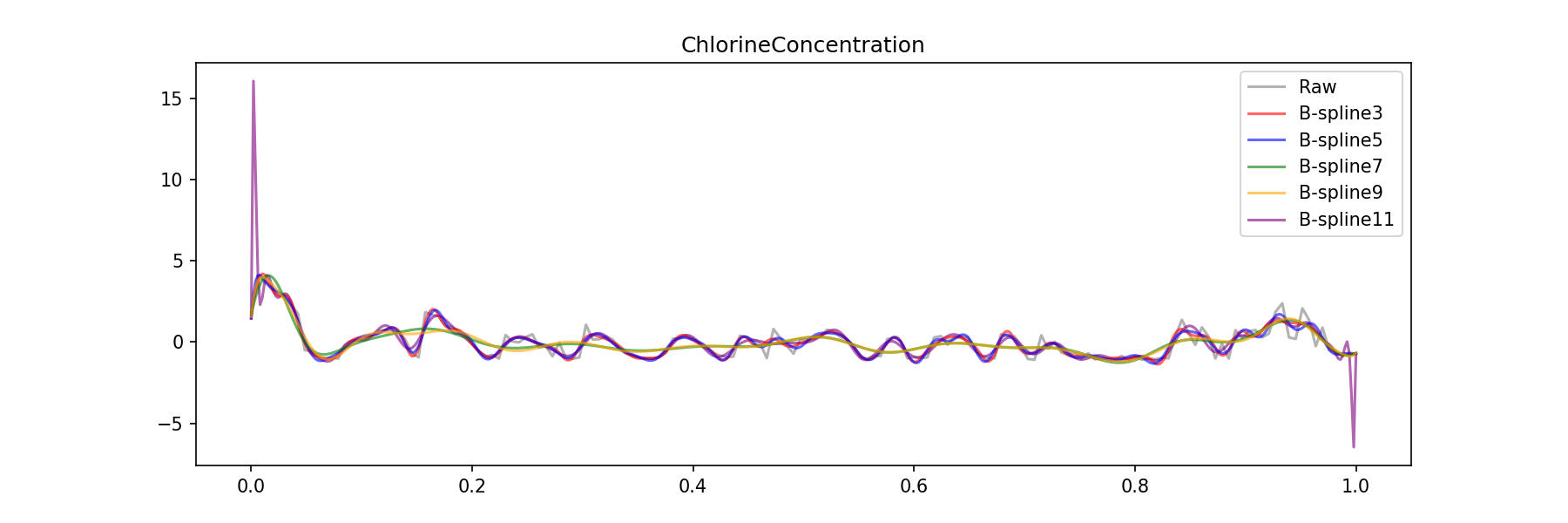}
    \caption{Different functional representations of the first instance in the ChlorineConcentration dataset.}
\end{figure}

\begin{figure}[!htb]
    \centering
    \includegraphics[width=1\linewidth]{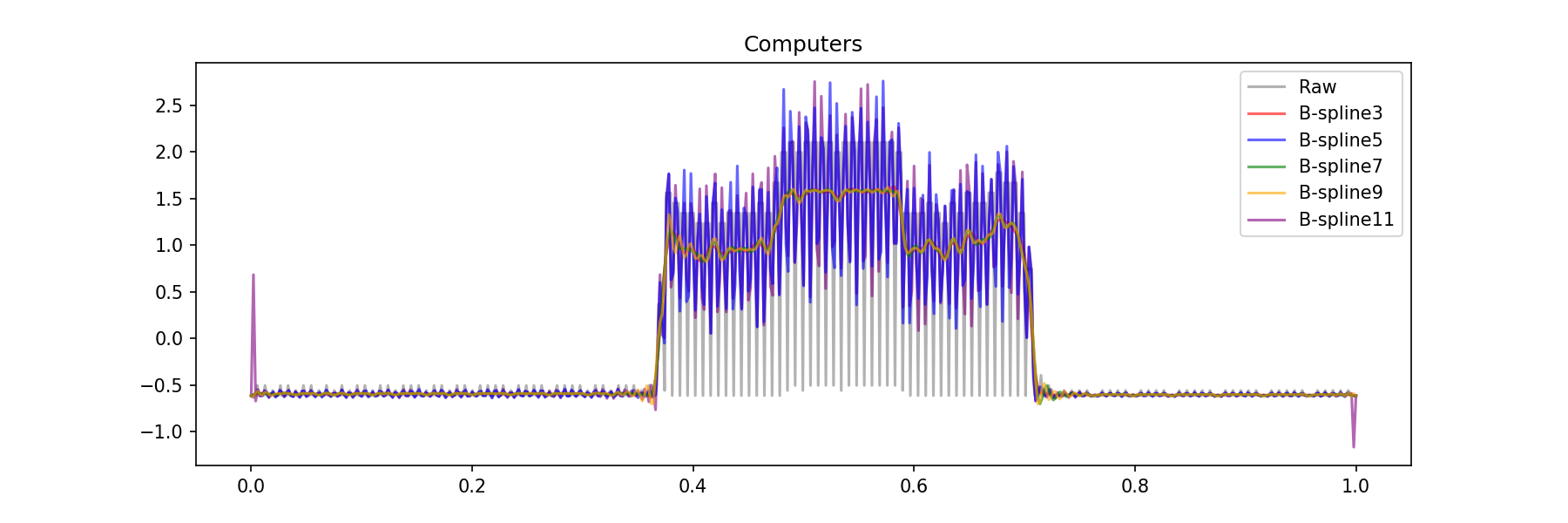}
    \caption{Different functional representations of the first instance in the Computers dataset.}
\end{figure}

\begin{figure}[!htb]
    \centering
    \includegraphics[width=1\linewidth]{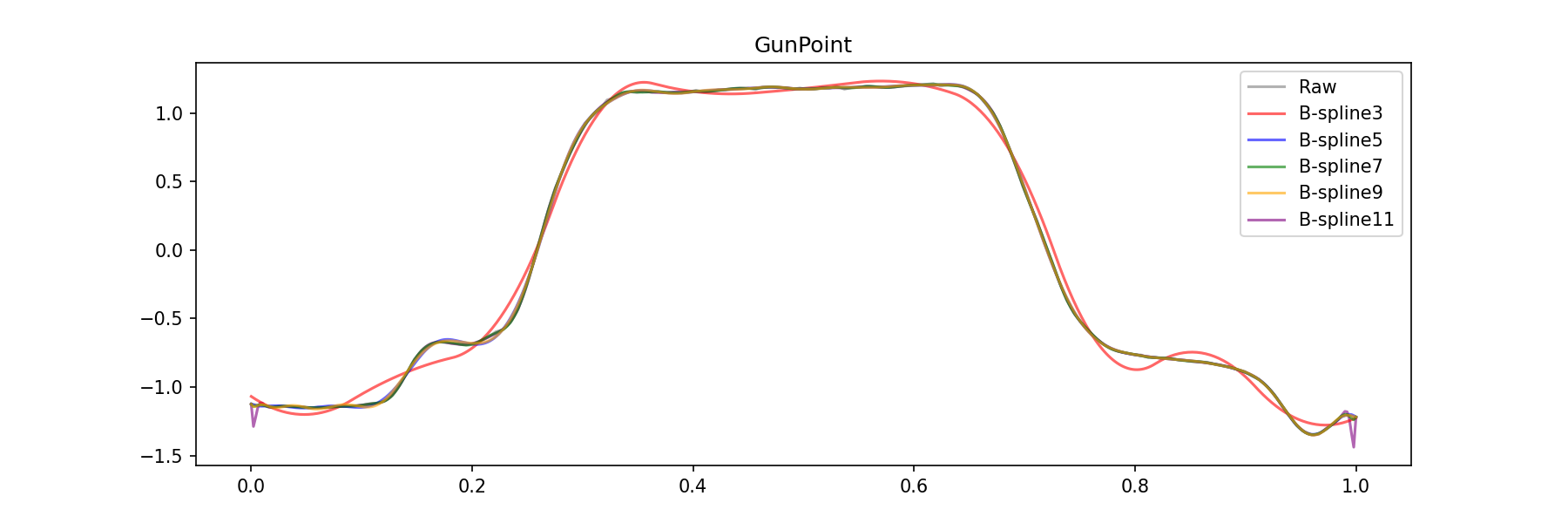}
    \caption{Different functional representations of the first instance in the GunPoint dataset.}
\end{figure}

\begin{figure}[!htb]
    \centering
    \includegraphics[width=1\linewidth]{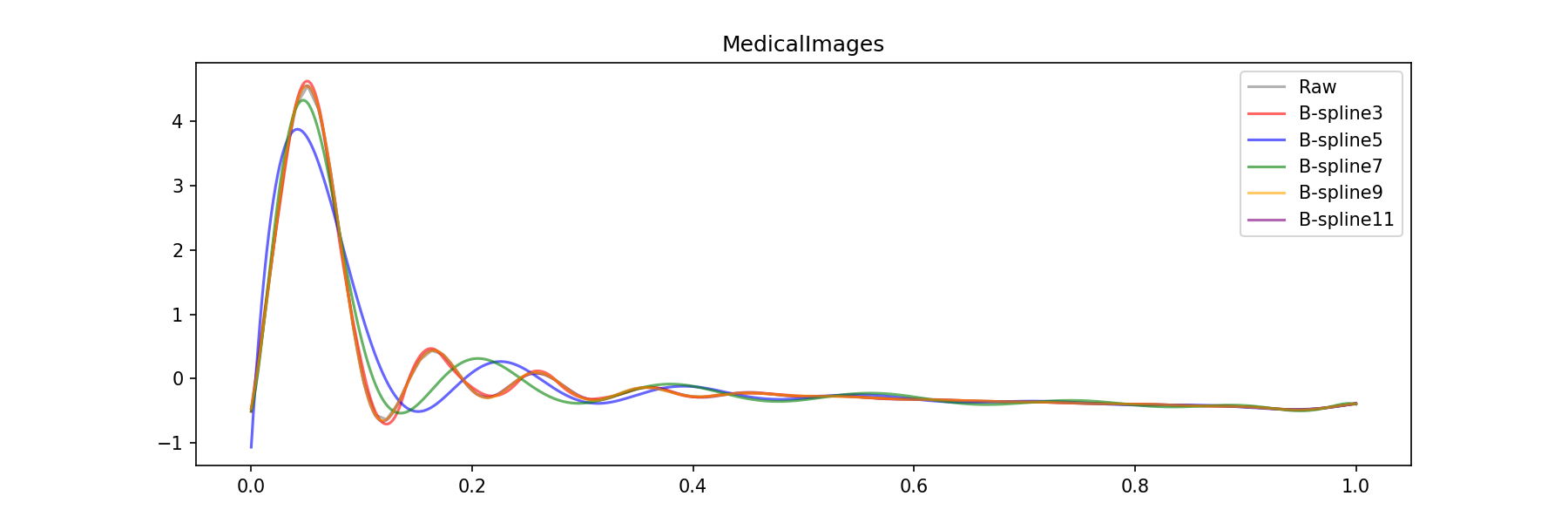}
    \caption{Different functional representations of the first instance in the MedicalImages dataset.}
\end{figure}

\begin{figure}[!htb]
    \centering
    \includegraphics[width=1\linewidth]{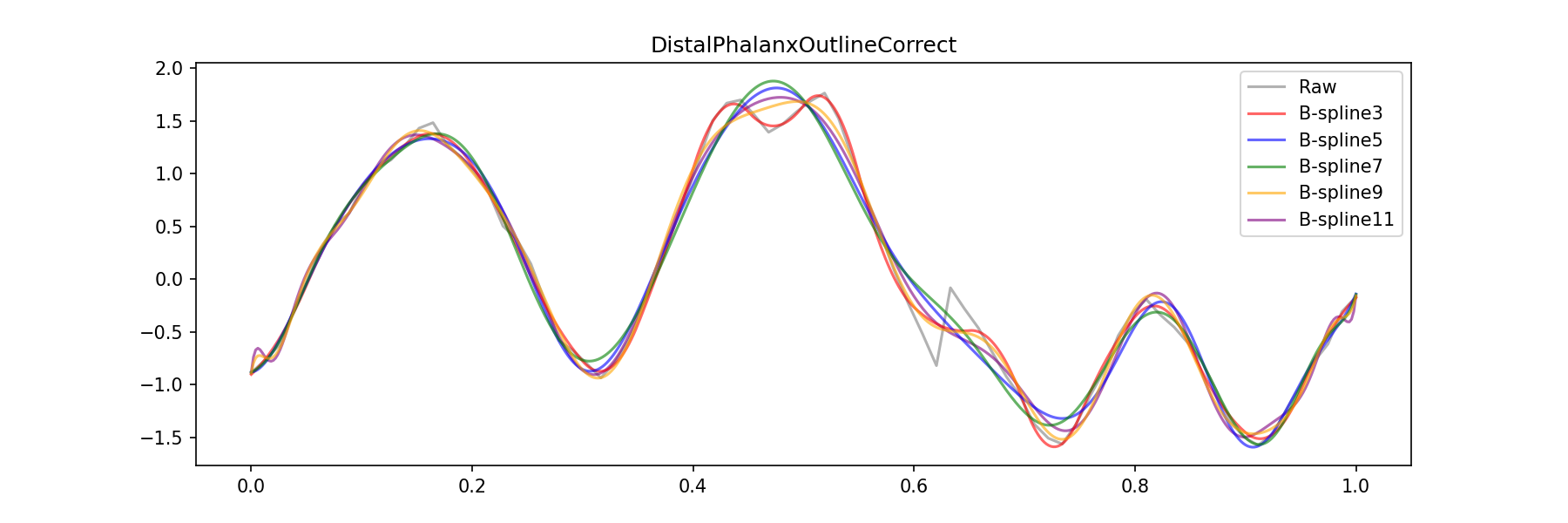}
    \caption{Different functional representations of the first instance in the DistalPhalanxOutlineCorrect dataset.}
\end{figure}

\begin{figure}[!htb]
    \centering
    \includegraphics[width=1\linewidth]{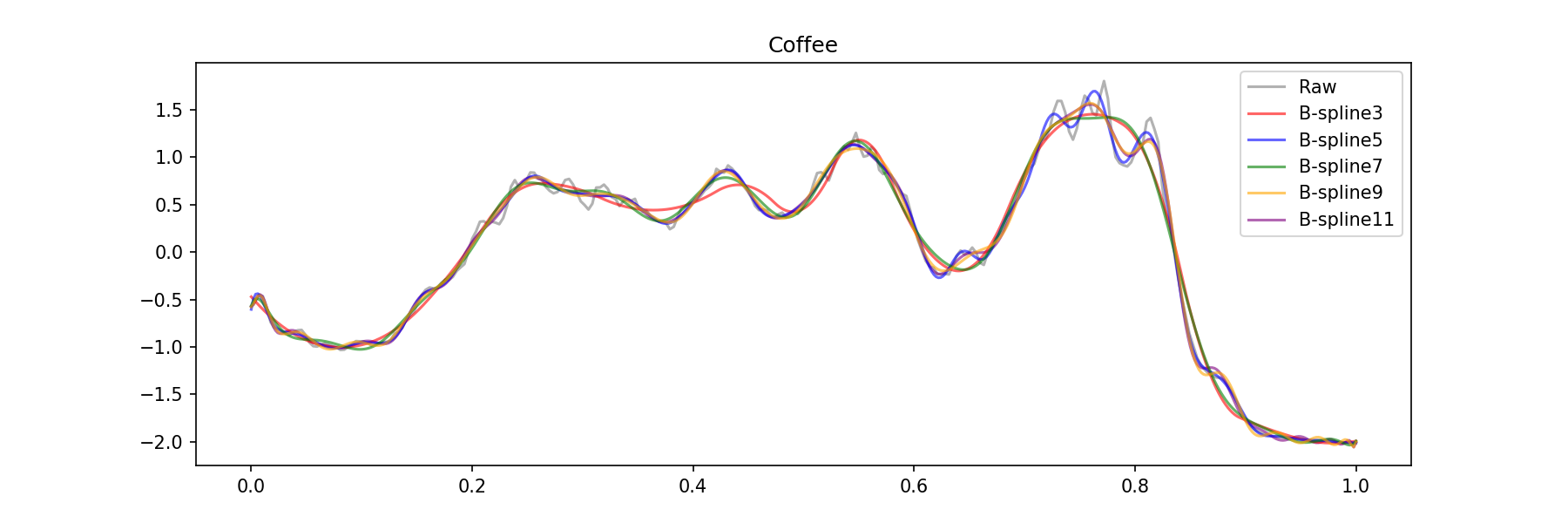}
    \caption{Different functional representations of the first instance in the Coffee dataset.}
\end{figure}

\begin{figure}[!htb]
    \centering
    \includegraphics[width=1\linewidth]{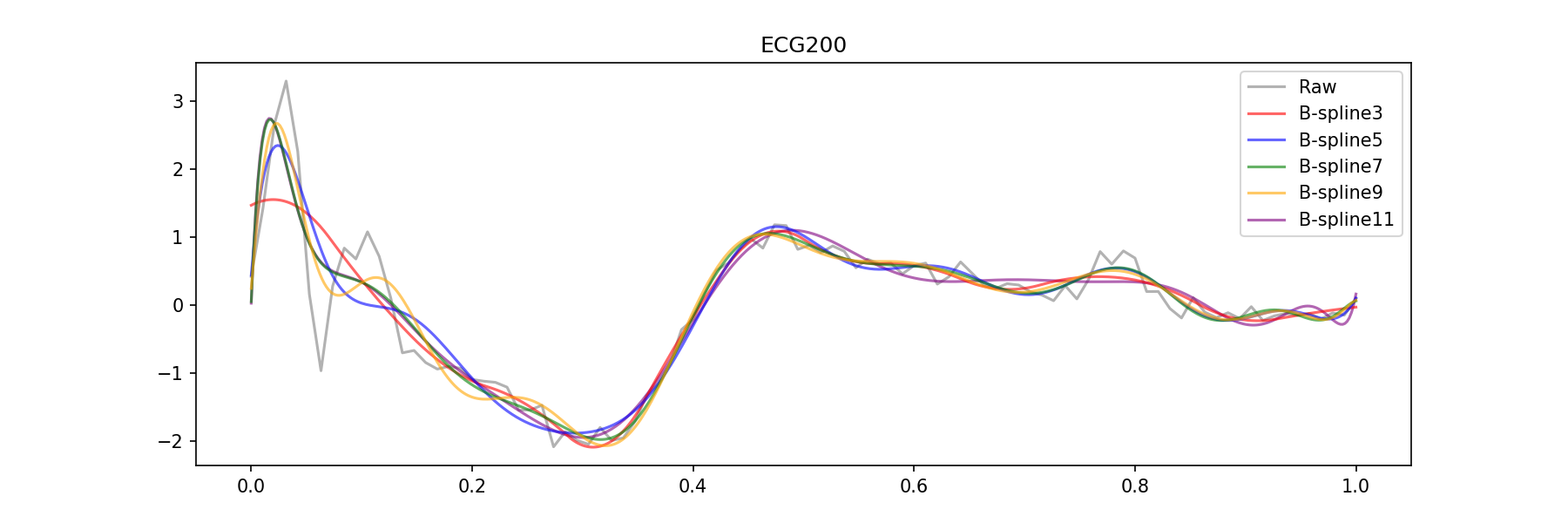}
    \caption{Different functional representations of the first instance in the ECG200 dataset.}
\end{figure}

\begin{figure}[!htb]
    \centering
    \includegraphics[width=1\linewidth]{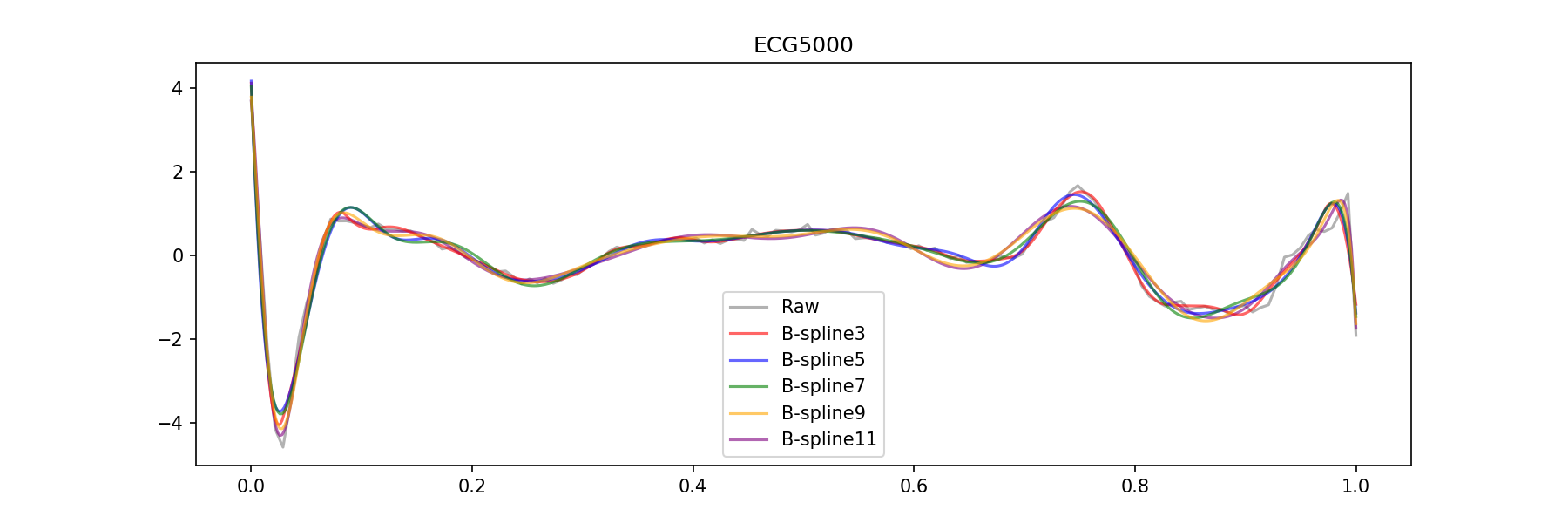}
    \caption{Different functional representations of the first instance in the ECG5000 dataset.}
\end{figure}

\begin{figure}[!htb]
    \centering
    \includegraphics[width=1\linewidth]{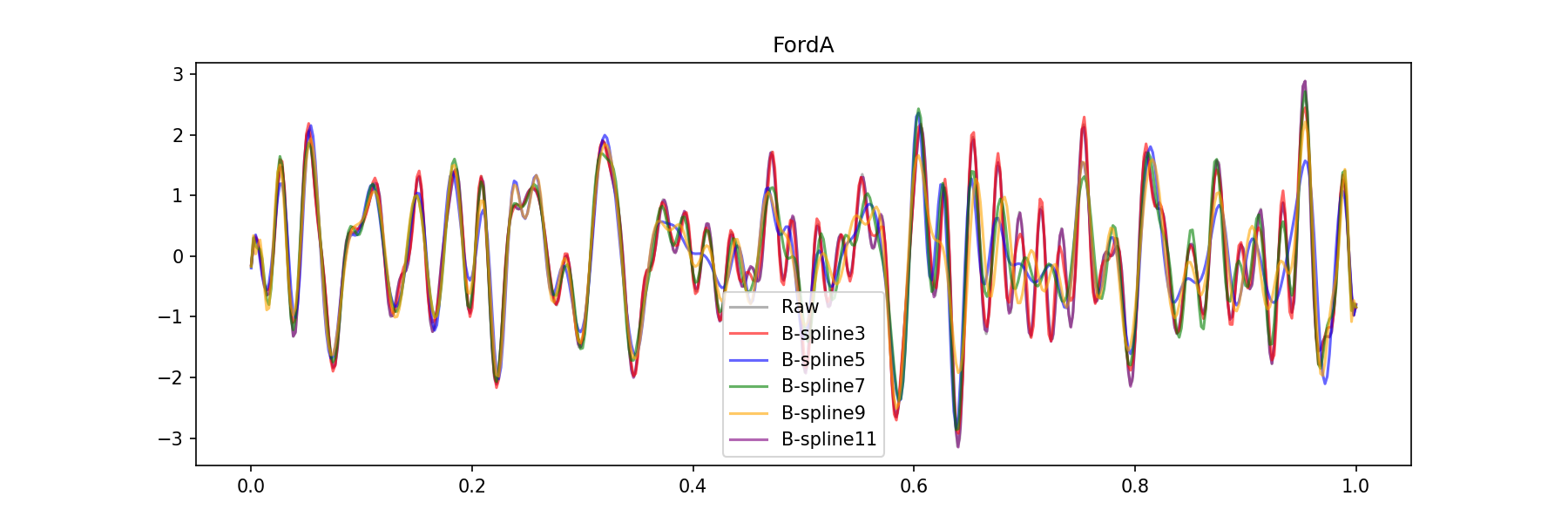}
    \caption{Different functional representations of the first instance in the FordA dataset.}
\end{figure}

\begin{figure}[!htb]
    \centering
    \includegraphics[width=1\linewidth]{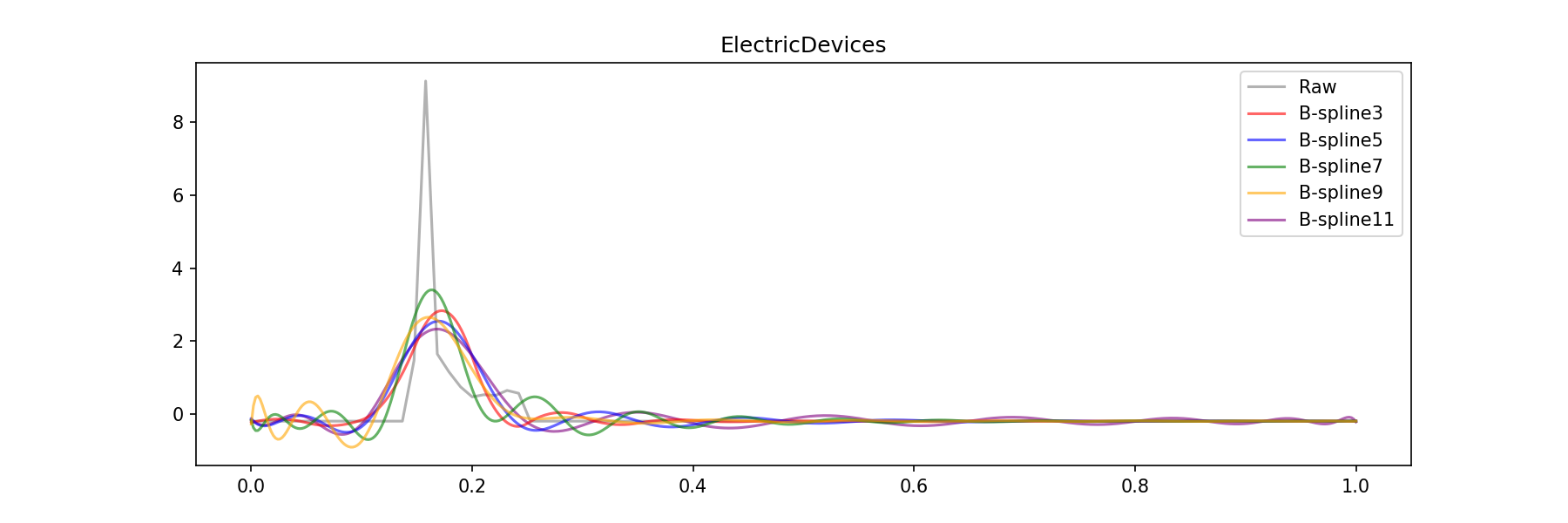}
    \caption{Different functional representations of the first instance in the ElectricDevices dataset.}
\end{figure}
\newpage
\clearpage

\bibstyle{spbasic}  
\bibliography{_biblio.bib}   

\end{document}